\documentclass[structabstract]{aa}
\usepackage{txfonts}
\usepackage{natbib}
\usepackage{graphicx}
\usepackage{color}
\usepackage{txfonts}
\bibpunct{(}{)}{;}{a}{}{,}

\title{Large-scale distributions of mid- and far-infrared emission from the center to the halo of M~82 revealed with AKARI}
\titlerunning{AKARI observations of M~82}
\authorrunning{H. Kaneda et al.}
\author{H. Kaneda\inst{1}, D. Ishihara\inst{1}, T.Suzuki\inst{2}, N. Ikeda\inst{3}, T. Onaka\inst{4}, M. Yamagishi\inst{1}, Y. Ohyama\inst{5}, T. Wada\inst{3}, \and A. Yasuda\inst{1} }
\institute{
Graduate School of Science, Nagoya University, Chikusa-ku, Nagoya 464-8602, Japan\\
\email{kaneda@u.phys.nagoya-u.ac.jp}
\and
Advanced Technology Center, National Astronomical Observatory of Japan, Mitaka, Tokyo 181-8588, Japan
\and
Institute of Space and Astronautical Science, Japan Aerospace Exploration Agency, Sagamihara, Kanagawa, 229-8510, Japan
\and
Graduate School of Science, The University of Tokyo, Bunkyo-ku, Tokyo 113-0033, Japan
\and
Academia Sinica, Institute of Astronomy and Astrophysics, Taipei 10617, Taiwan
}

\date{Received; accepted}

\abstract
{The edge-on starburst galaxy M~82 exhibits complicated distributions of gaseous materials in its halo, which include ionized superwinds driven by nuclear starbursts, neutral materials entrained by the superwinds, and large-scale neutral streamers probably caused by a past tidal interaction with M~81.}
{We investigate detailed distributions of dust grains and polycyclic aromatic hydrocarbons (PAHs) around M~82 to understand their interplay with the gaseous components.}
{We performed mid- (MIR) and far-infrared (FIR) observations of M~82 with the Infrared Camera and Far-Infrared Surveyor on board AKARI.}
{We obtain new MIR and FIR images of M~82, which reveal both faint extended emission in the halo and very bright emission in the center with signal dynamic ranges as large as five and three orders of magnitude for the MIR and FIR, respectively. We detect MIR and FIR emission in the regions far away from the disk of the galaxy, reflecting the presence of dust and PAHs in the halo of M~82.}
{We find that the dust and PAHs are contained in both ionized and neutral gas components, implying that they have been expelled into the halo of M~82 by both starbursts and galaxy interaction. In particular, we obtain a tight correlation between the PAH and H$\alpha$ emission, which provides evidence that the PAHs are well mixed in the ionized superwind gas and outflowing from the disk.}

\begin{document}
\maketitle

\section{Introduction}
M~82 is a nearby starburst galaxy in a group of galaxies, where an appreciable amount of material can be pushed out of a galaxy into the intergalactic medium by both internal (e.g. starburst activities) and external forces (e.g. tidal interactions between member galaxies). In fact, M~82 shows prominent galactic superwinds in H$\alpha$ \citep[e.g.][]{Bla88, Dev99} and X-rays \citep[e.g.][]{Bre95,Str97} accelerated out of the galactic plane, which are attributed to violent nuclear starbursts. The X-ray emission spatially correlates well with the H$\alpha$ emission \citep{Wat84,Str04}. The ionized galactic superwinds seem to entrain various phases of neutral gas (e.g. CO: Walter et al. 2002; H$_2$: Veilleux et al. 2009) and dust \citep{Alt99,Ohy02,Hoo05,Lee09}. In addition, M~82 shows large-scale molecular and atomic streamers anchoring around the edges of the galactic disk \citep{Wal02,Yun93}. The large-scale streamers extend mostly in parallel to the galactic plane and thus in entirely different directions from the superwinds. In particular, neutral hydrogen gas is largely extended around the intergalactic space of the M~81--M~82 group, including the halo regions of M~82 \citep{Yun94}. The presence of dust residing between the group members is also revealed through systematic reddening of photometric color of background galaxies viewed through the intergalactic medium \citep{Xil06}. The streamers are likely caused by the close encounter with M~81 that M~82 experienced about 100 Myr ago \citep{Yun93}. Then the hydrogen gas and dust in the intergalactic medium can be regarded as leftover from the interaction with M~81. 

As a result of its proximity ($\sim$3.5 Mpc) and nearly edge-on orientation with an inclination angle of about 80$^{\circ}$, M~82 is a valuable target for the study of extraplanar dust grains and their properties in its superwinds and galactic halos. To answer questions about how far, how much, and what kind of dust grains are carried out of the galaxy is of great importance for the understanding of material circulation and evolution in the galactic halo. The enrichment of the intergalactic medium with dust could affect observations of high-redshift objects \citep[e.g.][]{Hei88,Dav98}.  The dust expulsion from a galaxy would also play an important role in galactic chemical evolution acting as a sink for heavy elements \citep[e.g.][]{Eal96}. 

Nevertheless, information on the dust components in the halo of M~82 is relatively scarce in contrast to abundant information on the gaseous components. The dust in the halo was observed in reflection in the UV \citep{Hoo05} and optical \citep{Ohy02}, extinction \citep{Hec00}, and submillimeter dust continuum emission \citep{Lee09}. However, these are rather indirect or inefficient ways to detect largely-extended dust. High-sensitivity FIR observations from space are undoubtedly most effective to study the properties of faint extended emission from extraplanar dust, since dust emission typically peaks in the FIR and low photon backgrounds in space enable us to detect faint diffuse emission. However one serious problem is that the central starburst core is dazzlingly bright for space observations in the MIR and FIR, due to tremendous star-forming activity in the central region of M~82 \citep{Tel80}; M~82 is the brightest galaxy in the MIR and FIR after the Magellanic Clouds on the sky \citep{Col99}. Then, instrumental effects caused by saturation in observing very bright sources severely hamper reliable detection of low-level FIR emission outside the disk. The flux density of the central region of M~82 increases very rapidly towards wavelengths shorter than 300 $\mu$m; there is about two-orders-of-magnitude difference between FIR and submillimeter fluxes \citep{Thu00}, making FIR detection of the extraplanar dust in M~82 extremely difficult.

The situation in the MIR is similar for the very bright core, however, spatial resolution is much better in the MIR than in the FIR; the AKARI telescope of 700 mm in diameter has diffraction-limited imaging performance at a wavelength of 7 $\mu$m \citep{Kan07}. In addition to MIR dust continuum emission, star-forming galaxies show a series of strong MIR spectral features emitted by polycyclic aromatic hydrocarbons (PAHs) or PAH clusters \citep[e.g.][]{Smi07}, which can be regarded as the smallest forms of carbonaceous dust particles. The PAH emission features are unexceptionally bright in the central regions of M~82 \citep{Stu00,For03b}. With the Spitzer/IRAC and IRS, Engelbracht et al. (2006) and Beir\~ao et al. (2008) showed that the PAH emission is largely extended throughout the halo up to 6 kpc from the galactic plane. Their origins, again, can be either outflows entrained by the superwind or leftover clouds from the past interaction with M~81. Engelbracht et al. (2006) favored the latter origin because significant emission is detected outside the superwinds, and thus some process to expel PAHs from all parts of the disk is needed prior to the starburst. 

Far beyond the disk of the galaxy, extended emission named 'Cap' in the H$\alpha$ and X-ray was discovered at $\sim 11$ kpc to the north of the center of M~82 \citep{Dev99,Leh99}. The cap seen in the H$\alpha$ and X-ray may be the result of a collision between the hot superwind and a preexisting neutral cloud \citep{Leh99}. Hoopes et al. (2005) detected UV emission in the Cap suggesting that the emission is likely to be stellar UV light scattered by dust in the Cap. Tsuru et al. (2007) determined metal abundances of the X-ray plasma in the Cap region, which support the idea that the origin of the metal in the Cap is type-II supernova explosions that occurred in the central region of M~82. Hence the halo regions of M~82 are highly complicated; neither relationships among different phases of the outflowing material nor those between the superwinds and preexisting clouds is well understood.

In this paper, we report MIR and FIR imaging observations of M~82 performed with the Infrared Camera (IRC; Onaka et al. 2007) and the Far-Infrared Surveyor (FIS; Kawada et al. 2007), respectively, on board AKARI \citep{Mur07}. Thoughout this paper, we assume a distance of 3.53 Mpc for M~82 \citep{Kar02}. Our IRC data consist of 4 narrow-band images ($S7$, $S11$, $L15$, and $L24$) at reference wavelengths of 7 $\mu$m (effective band width: 1.75 $\mu$m), 11 $\mu$m (4.12 $\mu$m), 15 $\mu$m (5.98 $\mu$m), and 24 $\mu$m (5.34 $\mu$m), the allocation of which is ideal to discriminate between the PAH emission features ($S7$, $S11$) and the MIR dust continuum emission ($L15$, $L24$). The FIS has 4 photometric bands; 2 wide bands ($WIDE$-$S$ and $WIDE$-$L$) at central wavelengths of 90 $\mu$m (effective band width: 37.9 $\mu$m) and 140 $\mu$m (52.4 $\mu$m) and 2 narrow bands ($N60$ and $N160$) at 65 $\mu$m (21.7 $\mu$m) and 160 $\mu$m (34.1 $\mu$m). The wide bands provide high sensitivities, while the 2 narrow bands combined with the 2 wide bands are useful to accurately determine the temperatures of the FIR dust. Besides the fine allocation of the photometric bands, the special fast reset mode of the FIS as well as the combination of short and long exposure data of the IRC provide high signal saturation levels; we can safely observe very bright sources without serious saturation effects. Hence, the uniqueness of the IRC and FIS as compared to any other previous or currently existing instruments is a combination of their high saturation limits and high sensitivities (i.e. large dynamic range) with relatively high spatial resolution, which is essential to detect faint PAH and dust emission extending to the halo of an edge-on galaxy that is very bright in the center.

\section{Observations and data analyses}
We observed M~82 with AKARI seven times from April 2006 to April 2007. The observation log is listed in Table 1. The two observations of the IDs starting with '50' were performed during the AKARI performance verification phase. The observations with IDs starting with '51' were performed during AKARI Director's Time. The others were carried out in part of the AKARI mission program ``ISM in our Galaxy and Nearby galaxies'' (ISMGN; Kaneda et al. 2009a). 
In the two IRC observations with IDs starting with '51', we observed the Cap region, while the others were targeted at the galaxy body.

With the IRC, we obtained the $S7$, $S11$, $L15$, and $L24$ band images of M~82 using a standard staring observation mode, where each field-of-view has a size of about $10'\times 10'$. We also obtained the near-IR (NIR) $N3$ (reference wavelength of 3.2 $\mu$m) and $N4$ (4.1 $\mu$m) images simultaneously with the $S7$ and $S11$ images, but in this paper, we do not discuss the NIR images because there are no new findings different from the Spitzer/IRAC NIR images presented in Engelbracht et al. (2006). The MIR images were created by using the IRC imaging pipeline software version 20071017 (see IRC DATA User Manual; Lorente et al. 2007 for details). The background levels were estimated by averaging values from multiple apertures placed around the galaxy, while avoiding overlap with faint extended emission from the galaxy as much as possible, and were subtracted from the images.

The FIS observations were performed in the special fast reset mode called CDS (Correlated Double Sampling) mode in order to avoid signal saturation near the nuclear region of the galaxy. Fluxes were later cross-calibrated with the other ordinary integration modes by using the internal calibration lamp of the FIS. The FIR observations were performed three times with slightly shifted positions (table 1). A $10'\times 30'$ region was covered with every observation using a standard 2-round-trip slow scan mode. As a result, we covered an area of about $15'\times 15'$ for $N60$ and $WIDE$-$S$ and $20'\times 20'$ for $WIDE$-$L$ and $N160$ around M~82. The FIR images were processed from the FIS Time Series Data (TSD) using the AKARI official pipeline being developed by the AKARI data reduction team (Verdugo et al. 2007). The background levels were estimated from data taken near the beginning and the end of the slow scan observations and subtracted from the images.

To minimize detector artifacts due to the high surface brightness of the central starburst, we applied custom reduction procedures in addition to the above normal procedures. For the IRC, we combined the long exposure image with the short exposure image; the former has 28 times longer exposure than the latter, both obtained in one pointed observations. We replaced pixels significantly affected by the high surface brightness (i.e. saturated or even deviated from linearity) in the long exposure image by those unaffected in the short exposure image. In addition, very low-level ghost signals of the peak appeared at about $1'$ and $2'$ to the southwest direction of the center of M~82 for the $S7$ and $S11$ bands. We removed the ghosts by subtracting scaled images of the central $3'\times 3'$ area at the ghost positions for each band. 

For the FIS, low-level ghost signals appeared at about $5'$ to the north or south direction of the center, depending on whether the observation was performed in April or October. The reason for the ghost is electrical cross talk in the multiplexer of the cryogenic readout electronics \citep{Kaw07}; the narrow band produces a ghost signal just at the timing when the wide band detects a strong signal, and vice versa. We removed the ghost signals by masking the TSD where the ghosts are predicted to appear before creating the images. We also removed residual artifacts due to cosmic-ray hits by masking the affected TSD. The lost data were replaced by utilizing data redundancy gained from the three observations. Note that we did not apply any high-pass filtering because the high-pass filtering not only removes streaking due to residual slow response variations of the detector but also filters out faint emission extended around the galaxy. Instead, only for $N160$ where the streaking was relatively strong and larger blank areas were obtained, the residual variations were corrected by using blank sky for flat-fielding in both upstream and downstream regions of the slow scan. 
 
\begin{table*}
\caption{Observation log}
\label{log}
\centering
\renewcommand{\footnoterule}{}
\begin{tabular}{cccccc}
\hline\hline
Instrument & R.A. (J2000) & Dec. (J2000) & Observation ID & Date \\
\hline
FIS & 09 55 33.0 & $+$69 41 55.3 & 5011011 & 2006 Apr 18 \\
FIS & 09 56 11.3 & $+$69 39 42.1 & 5011012 & 2006 Apr 18 \\
IRC & 09 55 52.2 & $+$69 40 46.9 & 1400586 & 2006 Oct 21 \\
FIS & 09 55 52.2 & $+$69 40 46.9 & 5110031 & 2006 Oct 22 \\
IRC & 09 55 26.0 & $+$69 48 48.6 & 5124075 & 2007 Apr 19 \\
IRC & 09 55 26.0 & $+$69 48 48.6 & 5124076 & 2007 Apr 19 \\
IRC & 09 55 52.2 & $+$69 40 46.9 & 1401057 & 2007 Apr 20 \\
\hline
\end{tabular}
\end{table*}

\section{Results}
\subsection{MIR images}
The MIR 4-band contour images of the central $8'\times 8'$ area of M~82 obtained with the IRC are shown in Fig.1, where the bin size for all the maps is set to be 2\farcs3. The FWHM of the point spread function (PSF) is about $5''$ for each band \citep{Ona07}. These maps exhibit distributions of surface brightness over a range as large as 4 orders of magnitude, demonstrating the large dynamic range of the IRC. The peak surface brightness is 5130, 4341, 18344, and 31842 MJy sr$^{-1}$ for $S7$, $S11$, $L15$, and $L24$, respectively. The 1-sigma background fluctuation levels are approximately 0.09, 0.06, 0.1, and 0.9 MJy sr$^{-1}$, which correspond to $3-14$ \% of the lowest contour levels in the MIR maps. 

Based on the Spitzer/IRS spectroscopy of the central region of M~82 \citep{Bei08}, the $S7$ and $S11$ band images are most likely dominated by the PAH emission features, while the $L15$ and $L24$ images are dominated by hot dust continuum emission by very small grains (VSGs). On the basis of the observed spectrum combined with the IRC system response curve \citep{Ona07}, we estimate that PAH emission contributes approximately 85 \% and 45 \% of the band intensities of $S7$ and $S11$, respectively near the central region, and even higher in the halo because hot continuum emission is much less likely in regions far from the central starburst. This is supported by the fact that the $L24$ band image shows the most compact distribution, while the $L15$ band image shows less extended structures but may contain a small contribution from the PAH 17 $\mu$m broad emission feature \citep{Wer04,Kan08b}. 
 
The bright emission in the central part extends more or less along the east-west direction for each map, corresponding to the major axis of the M~82 optical disk. The very central region exhibits a double peak structure, where the stronger peak does not coincide with the optical center of M~82, slightly shifting to the west. A similar double-lobed distribution of the inner disk was also observed in CO emission \citep{She95} and submillimeter dust continuum emission \citep{Lee09}. At larger scales, all the maps show extended emission structure in the northwest and southeast directions, although the $L24$ map apparently suffers from the diffraction spike pattern of the telescope truss.
% The $S7$ and $S11$ images show more extended emission than the $L15$ and $L24$ images.

There is a striking similarity between the $S7$ and $S11$ band images; there are two filamentary structures in the northern halo and one in the southern halo. The similarity is well consistent with dominance of the PAH emission in both bands. The northern two filaments are of similar brightness in the $S7$ and $S11$ images, while only the western structure is clearly seen in the $L15$ and $L24$ band images. This implies a difference in spatial distribution between the PAHs and VSG emission in the two extended structures, presumably a difference not in material distribution but in radiation field intensity (see section 4.1). The southern filament is seen also in the $L15$ image, but not clearly in the $L24$ image due to the diffraction pattern. The central 2\farcm5 maps of the 850 $\mu$m continuum emission \citep{Lee09} and integrated CO(2$-$1) line intensity \citep{Thu00} exhibit only the southern extended structure.  

\begin{figure*}
\includegraphics[width=0.5\textwidth]{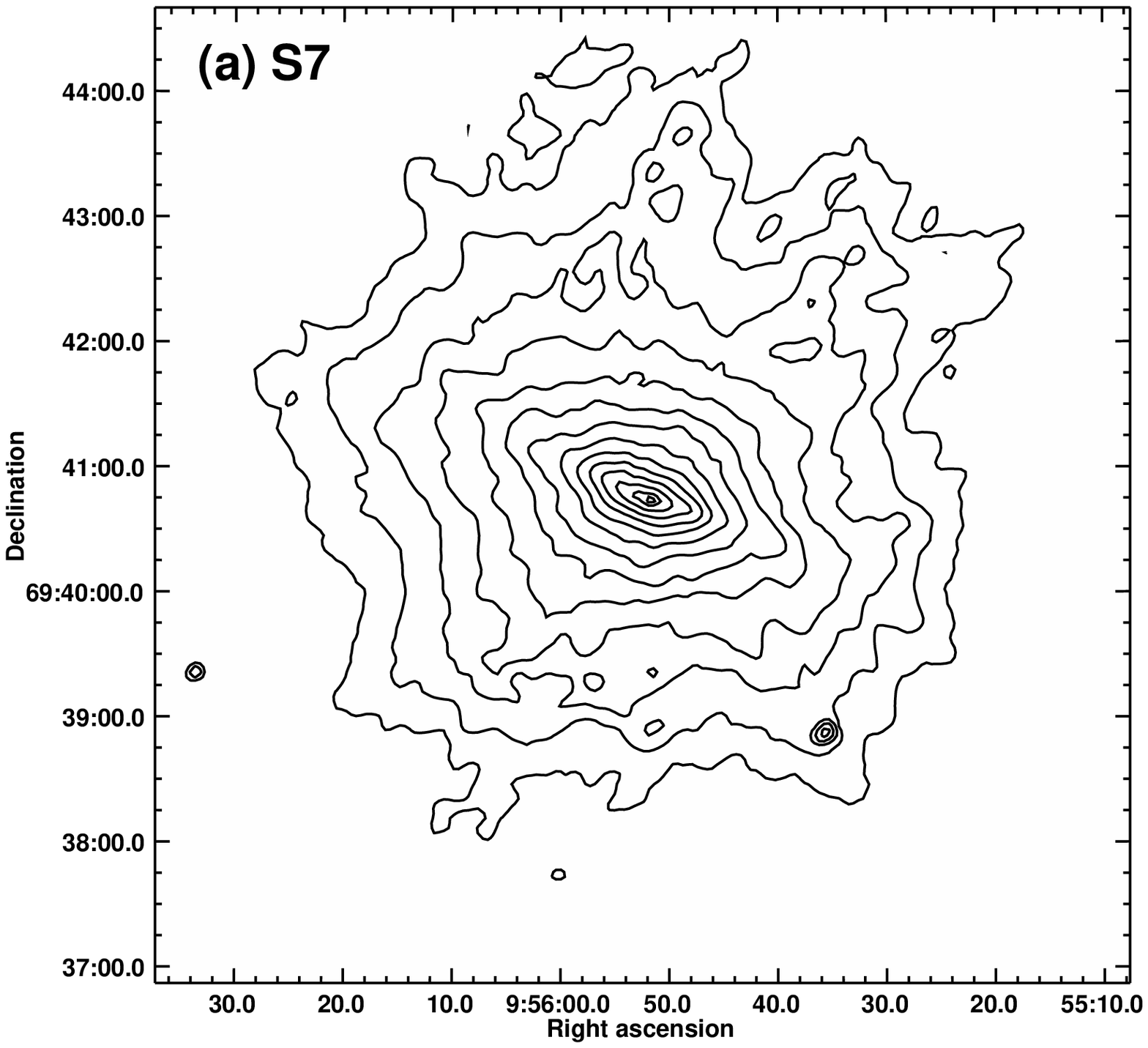}
\includegraphics[width=0.5\textwidth]{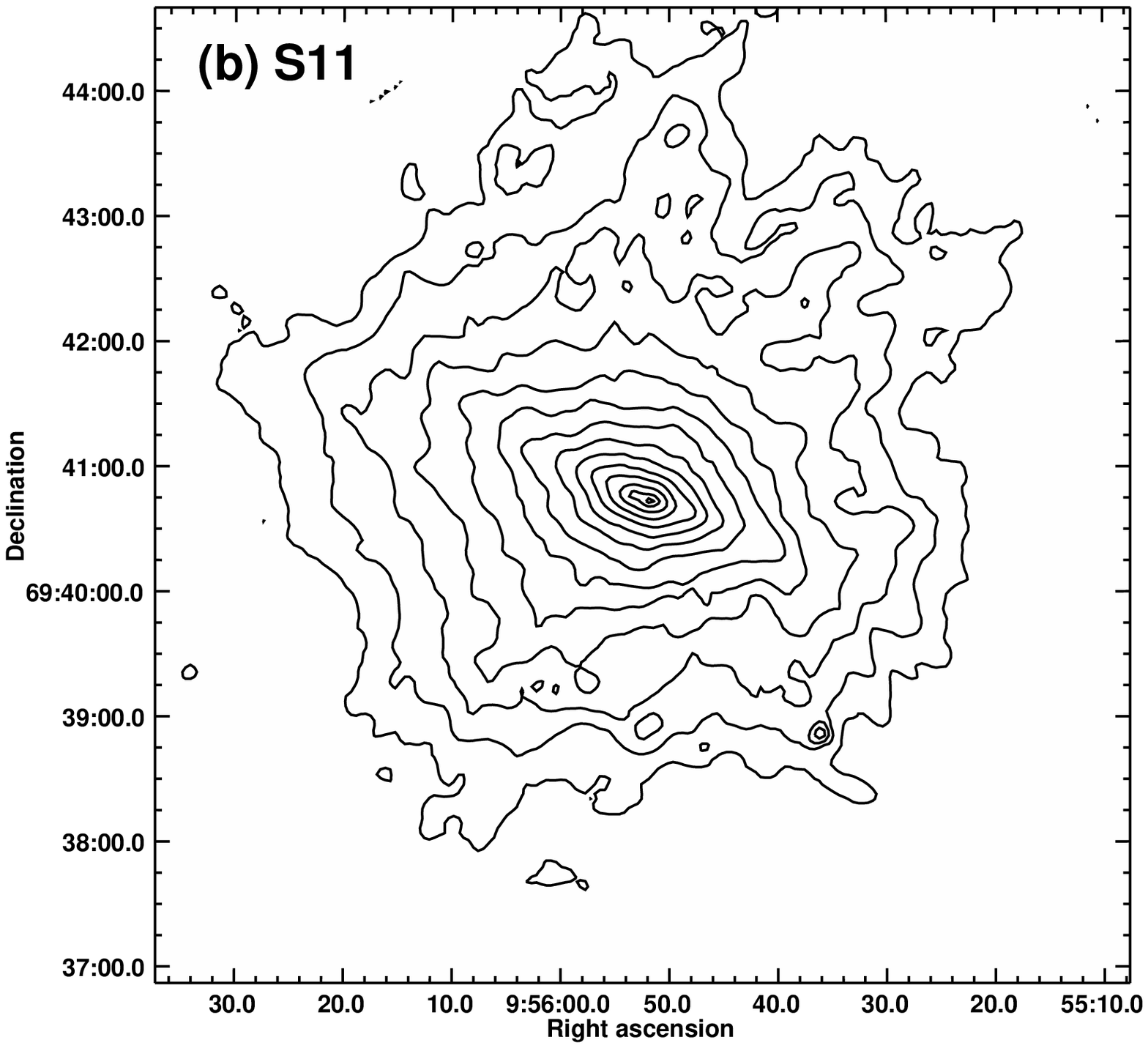}\\
\includegraphics[width=0.5\textwidth]{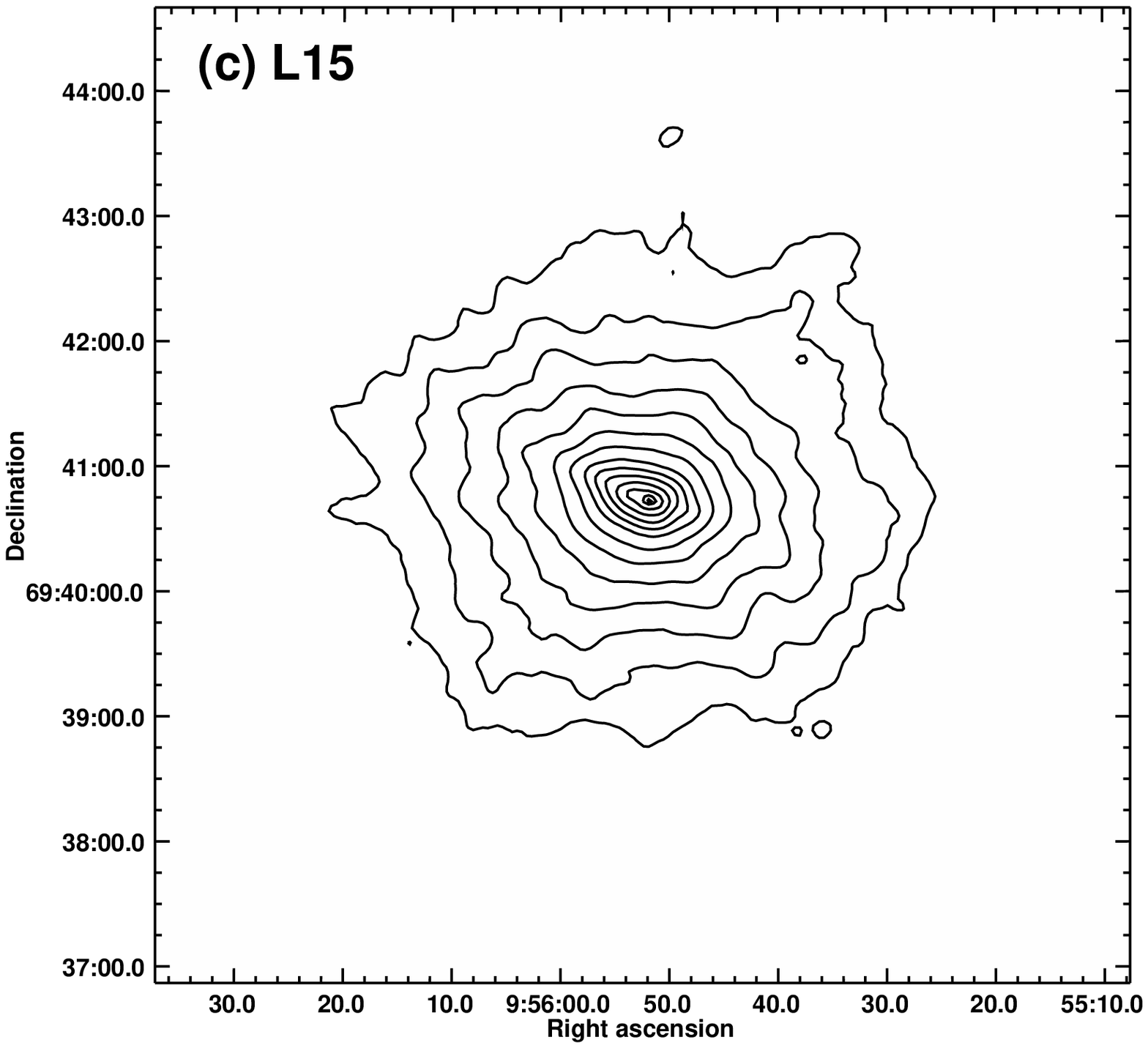}
\includegraphics[width=0.5\textwidth]{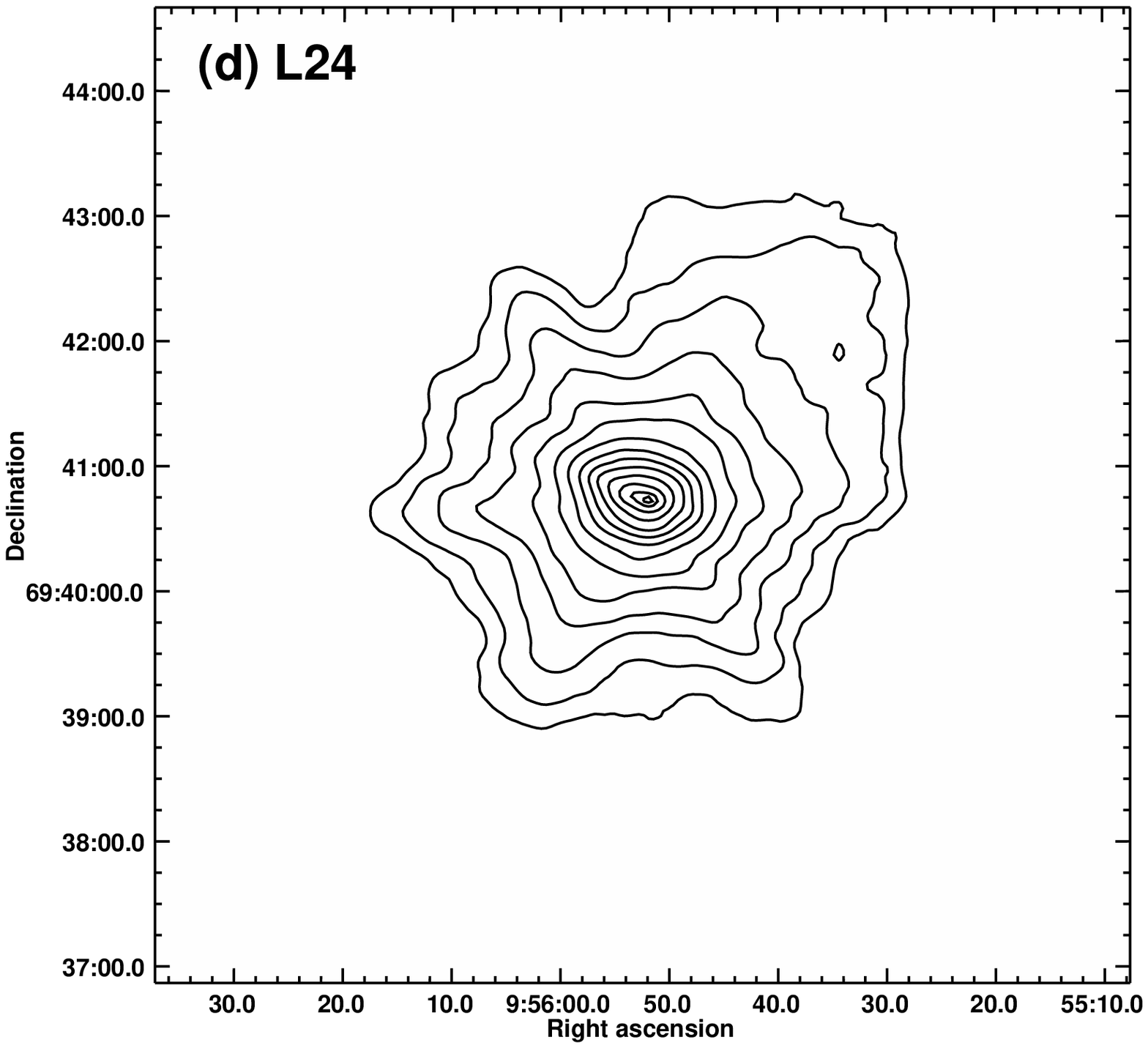}
\caption{MIR images of M~82 obtained with the AKARI/IRC in the (a) $S7$, (b) $S11$, (c) $L15$, and (d) $L24$ bands at the reference wavelengths of 7, 11, 15, and 24 $\mu$m, respectively. The contours are drawn at 95 \% and logarithmically-spaced 13 levels from 80 \% to 0.02 \% of the peak surface brightness, 5130, 4341, 18344, and 31842 MJy/str for $S7$, $S11$, $L15$, and $L24$, respectively.}

\end{figure*}

\subsection{FIR images}
The FIR 4-band contour images of the central $40'\times 40'$ area of M~82 obtained with the FIS are shown in Fig.2, where the bin size for all the maps is set to be $15''$. The FWHM of the PSF is $37''-39''$ for $N60$ and $WIDE$-$S$ and $58''-61''$ for $WIDE$-$L$ and $N160$ \citep{Kaw07}. The grey-scale image from the same data is also shown in order to reveal spatial coverage for each band. Note that the observed area is slightly different from band to band; $N160$ covers the northernmost and westernmost area, while $WIDE$-$L$ covers the largest area among the bands. The peak surface brightness is 9881, 7569, 5974, and 4195 MJy sr$^{-1}$ for $N60$, $WIDE$-$S$, $WIDE$-$L$, and $N160$, respectively. The 1-sigma background fluctuation levels are 2.0, 0.5, 1.0, and 1.5 MJy sr$^{-1}$ for $N60$, $WIDE$-$S$, $WIDE$-$L$, and $N160$, respectively. They correspond to $7-18$ \% levels of the lowest contours that are as low as $0.1-0.2$ \% levels of the peak brightness. The elongation in the east-west direction seen in the $N60$ and $WIDE$-$S$ images is a result of optical cross talk among the pixels of the monolithic arrays used for these bands \citep{Kaw07}. Telescope diffraction patterns similar to that in the $L24$ image (Fig.1d) are also visible in the $N60$ and $WIDE$-$S$ images. The elongation along the scan direction seen in the $WIDE$-$L$ image is probably due to the slow transient response of the detector used for this band. Other than these, we believe there are no additional substantial artifacts in the FIR images. Note that the spatial scale in Fig.2 is much larger than that in Fig.1; the MIR image size in Fig.1 is indicated by the dashed square in Fig.2d. Alton et al. (1999) stressed that the central FIR emission should be very compact juding from the compactness of the submillimeter emission, which is consistent with our results. It should be noted that the intensity scales in Fig.2 are stretched in order to emphasize the low surface brightness emission in the halo.

The emission component extending to the northwest direction from the center can be recognized in the $WIDE$-$S$, $WIDE$-$L$, and $N160$ images, and is the most prominent in the $N160$ band because the extended emission is cooler than the central emission. In addition, there are two emission regions away from the galaxy body clearly seen in both $WIDE$-$S$ and $WIDE$-$L$ images, one located at $12'$ to the west and the other at $8'$ to the southeast from the center. There are no counterparts found in the SIMBAD database. For the southeast emission region that is located inside the borders of the observed area in both bands, the ratio of the $WIDE$-$S$ to the $WIDE$-$L$ brightness corresponds to a color temperature of 23 K, which is significantly higher than typical temperatures of $16-18$ K for the Galactic cirrus \citep{Sod94} and thus unlikely of Galactic foreground origin. As shown later, these emission structures seem to be spatially correlated with the neutral gas streamers. There is another emission region seen only in the $N60$ band, at about $8'$ to the northwest from the center, which is located approximately on the line extending toward the direction of the above elongated structure seen in the $WIDE$-$S$, $WIDE$-$L$, and $N160$ images. The good spatial alignment might indicate that the source is associated with the elongated structure. The contribution of gas emission lines such as the [\ion{O}{i}] 63 $\mu$m line could be very important because this line can dominate the flux of the IRAS 60 $\mu$m band, similar to the $N60$ band, in shocked regions \citep{Bur90}. However the previous mapping of M~82 by shock tracers such as SiO \citep{Gar01}, [FeII] 1.644 $\mu$m \citep{Alo03}, and molecular hydrogen \citep{Vei09} does not cover the above bright region that is located too far ($\sim 8'$) from the nucleus.

\begin{figure*}
\includegraphics[width=0.5\textwidth]{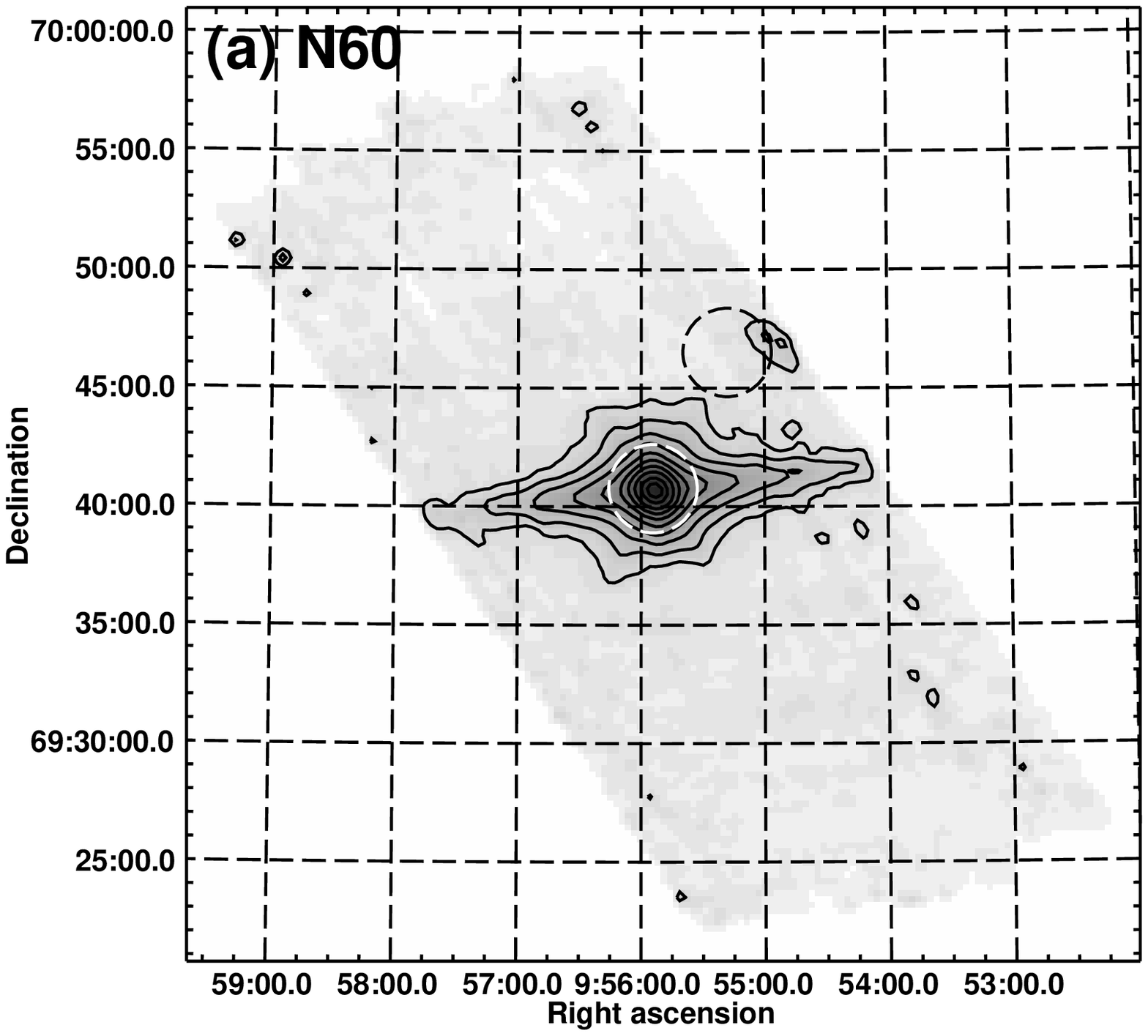}
\includegraphics[width=0.5\textwidth]{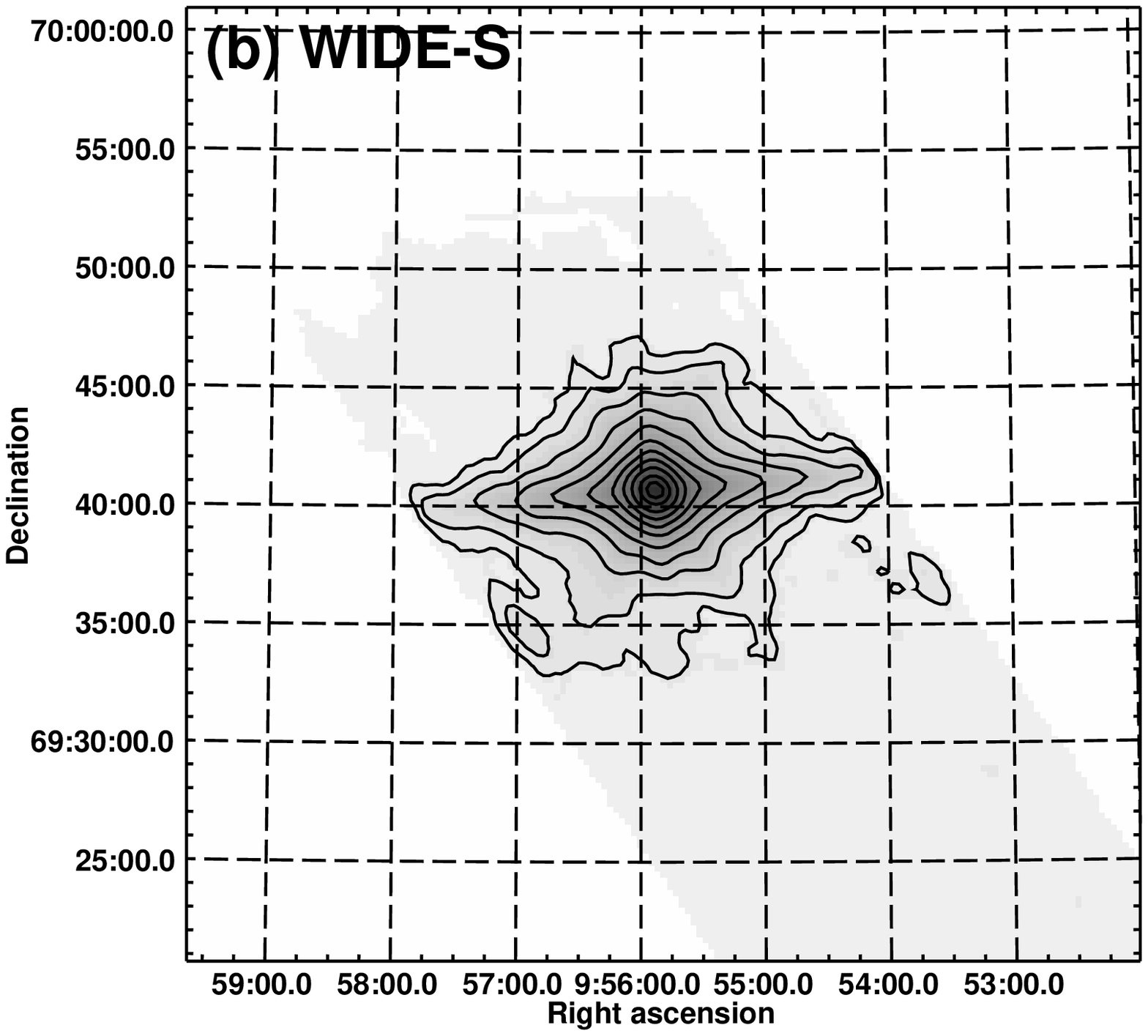}\\
\includegraphics[width=0.5\textwidth]{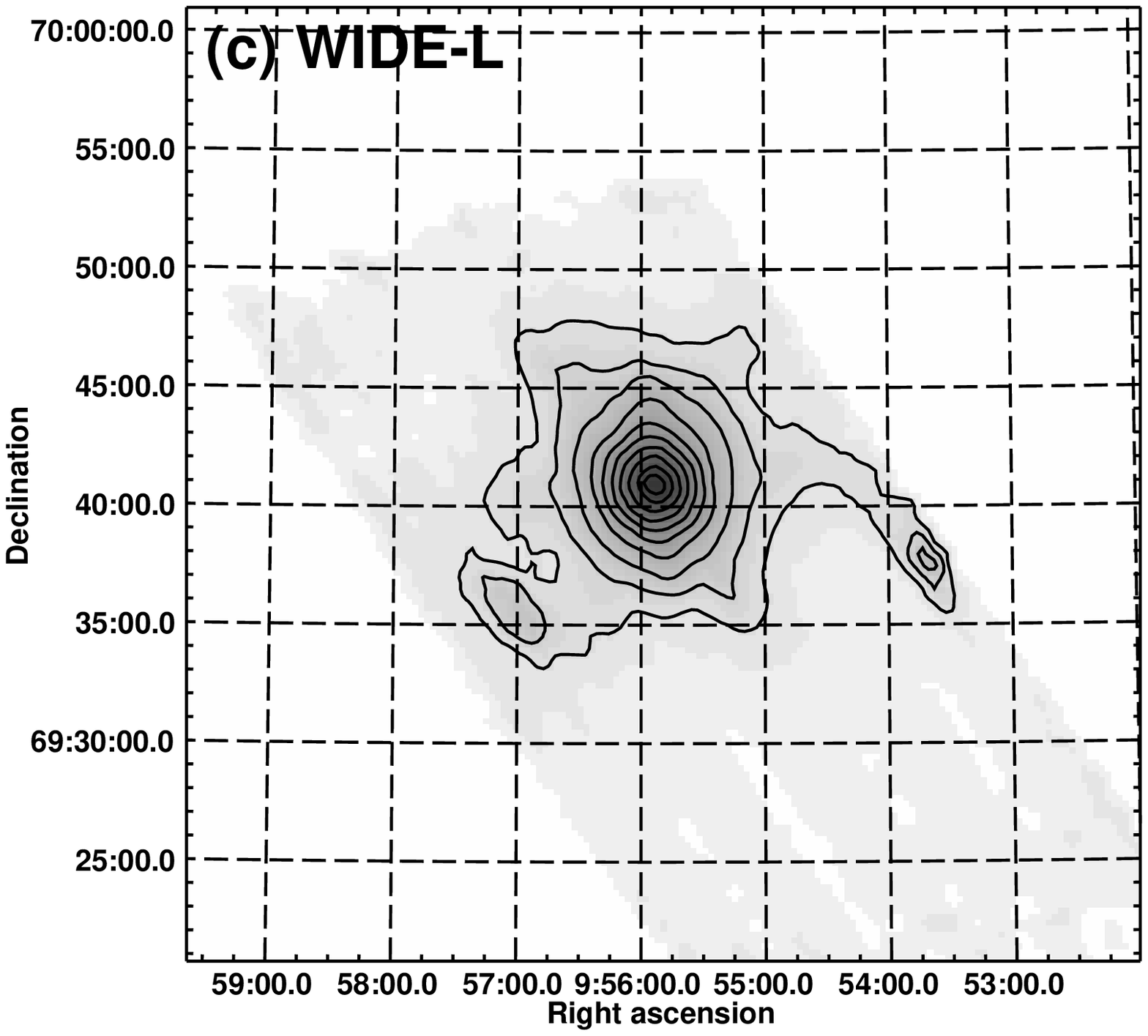}
\includegraphics[width=0.5\textwidth]{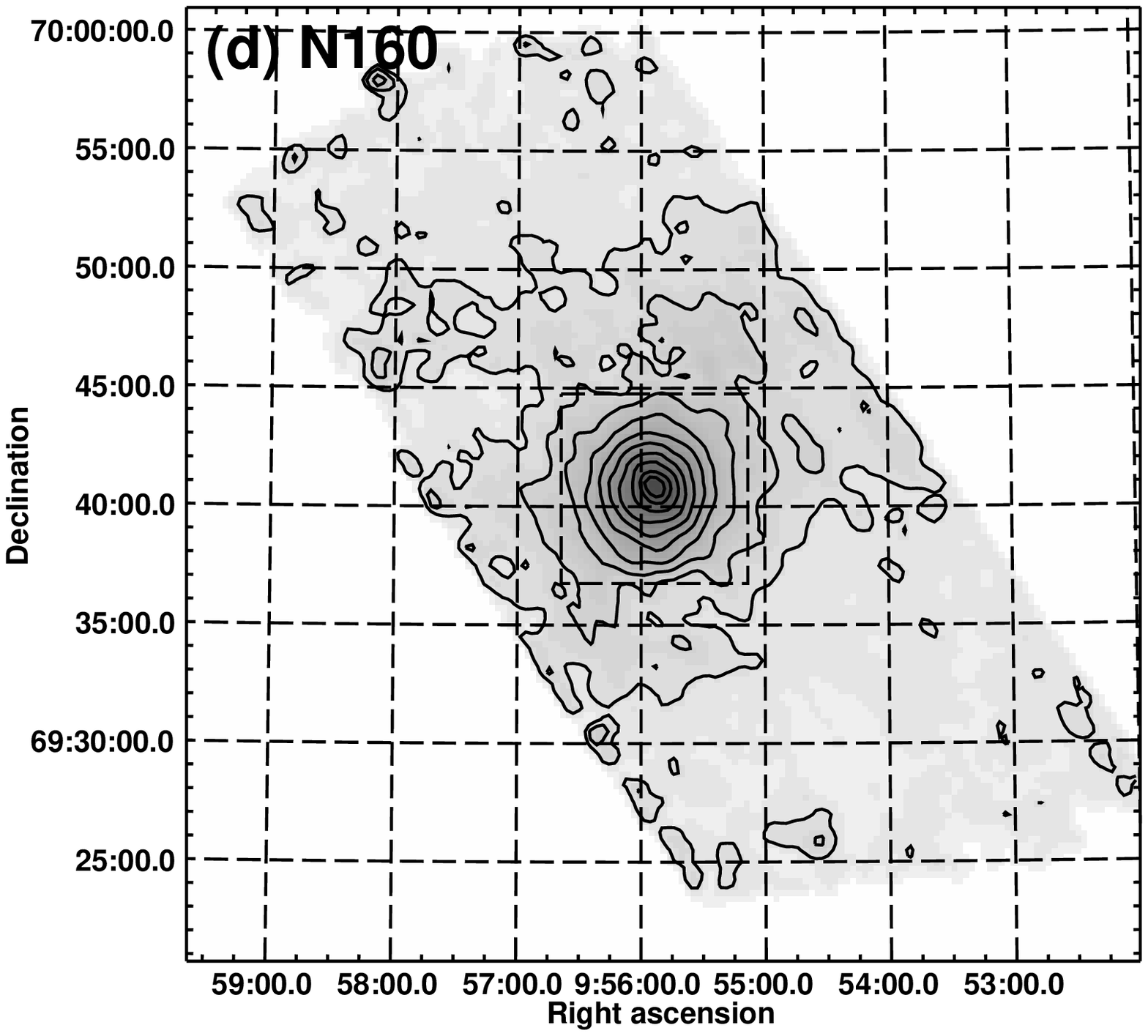}
\caption{FIR images of M~82 obtained with the AKARI/FIS in the (a) $N60$, (b) $WIDE$-$S$, (c) $WIDE$-$L$, and (d) $N160$ bands at the central wavelengths of 65, 90, 140, and 160 $\mu$m, respectively. The contours are drawn at logarithmically-spaced 10 levels from 80 \% to 0.2 \% of the peak surface brightness, 9881, 7569, 5974, and 4195 MJy/str for $N60$, $WIDE$-$S$, $WIDE$-$L$, and $N160$, respectively. Only for $WIDE$-$S$, the 0.1 \% level contour is added. Note that the elongation of the central emission along the east-west direction in $N60$ and $WIDE$-$S$ as well as that along the scan direction in $WIDE$-$L$ is a result of detector artifacts (see text). The two circular apertures for photometry of the center (white) and halo (black) regions ($d\leq 4'$) are shown in the panel (a), while the MIR image size in Fig.1 is indicated by the dashed square in the panel (d).}

\end{figure*}

\subsection{Cap region}
With the IRC, we performed dedicated observations of the halo region including the X-ray Cap located at $\sim 11'$ to the north from the center of M~82. Figures 3a$-$d show the obtained images, where we combine the two $10'\times 10'$ fields-of-view of the IRC, one centered at the galaxy body and the other at the Cap region. The color scales are stretched to very low surface brightness levels to bring out faint diffuse emission in the halo, and therefore the disk of the galaxy that is presented in Fig.1 is saturated in Fig.3. For comparison, in Fig.3e, we show the FIR $N160$ band image of the same area as in Fig.3a$-$d; among the 4 FIR bands, only the $N160$ band covers the X-ray Cap. In creating the MIR images in Fig.3a$-$d, we slightly shift the dark levels of the images of the Cap region so that it can be connected smoothly to the images of the galaxy body. We have applied smoothing to the MIR images by a gaussian kernel of $24''$ in FWHM to increase S/N ratios for detecting faint diffuse emission. 

There seems to be a faint emission component extended largely in the Cap region, especially in the $S9$ and $S11$ band images. The color scales of these images are logarithmically scaled down to 0.0007 \% of the peak surface brightness of the central starburst, which are still approximately 6, 5, 7, and 6 times higher than the 1-sigma background fluctuation levels in the corresponding bands. Here, since there might be some offset in dark levels between the two fields-of-view, we re-estimate the background level and its fluctuation within a single aperture located at the darkest nearby sky of the smoothed image as shown in Fig.3d. Hence the presence of the largely-extended faint emission component is statistically significant. The FIR image shows the aforementioned emission extended toward the northwest direction, but no significant signal at the position of the Cap. 

\begin{figure*}
\includegraphics[width=0.33\textwidth]{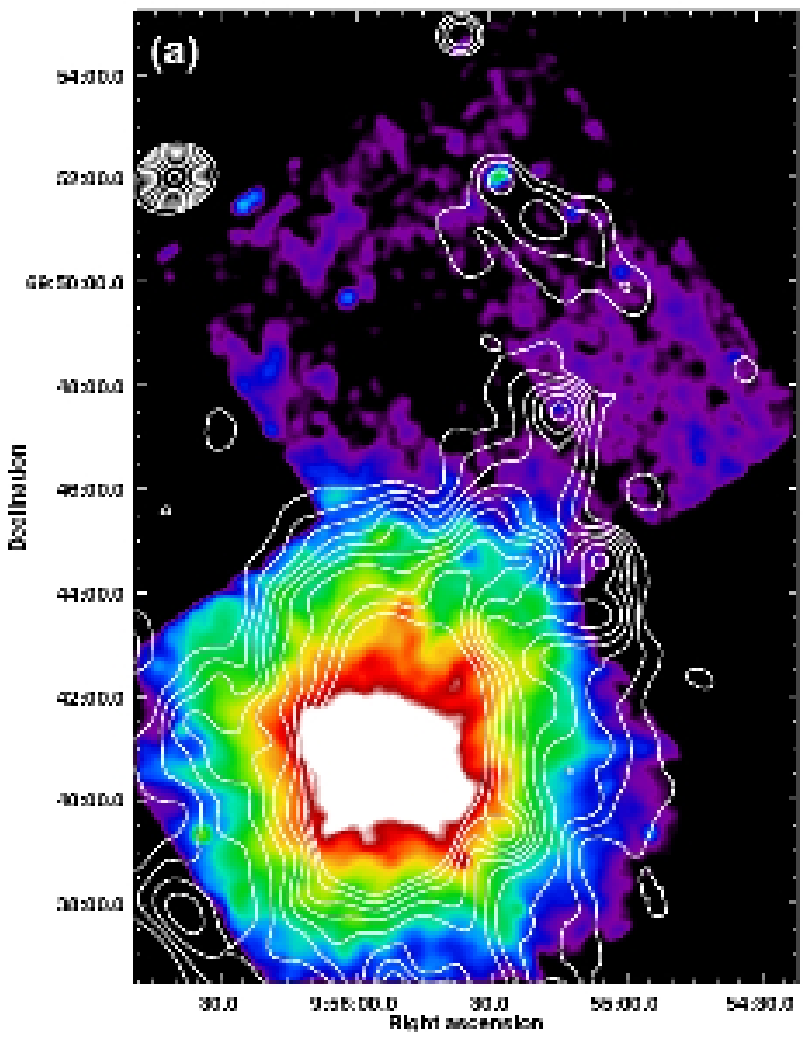}
\includegraphics[width=0.33\textwidth]{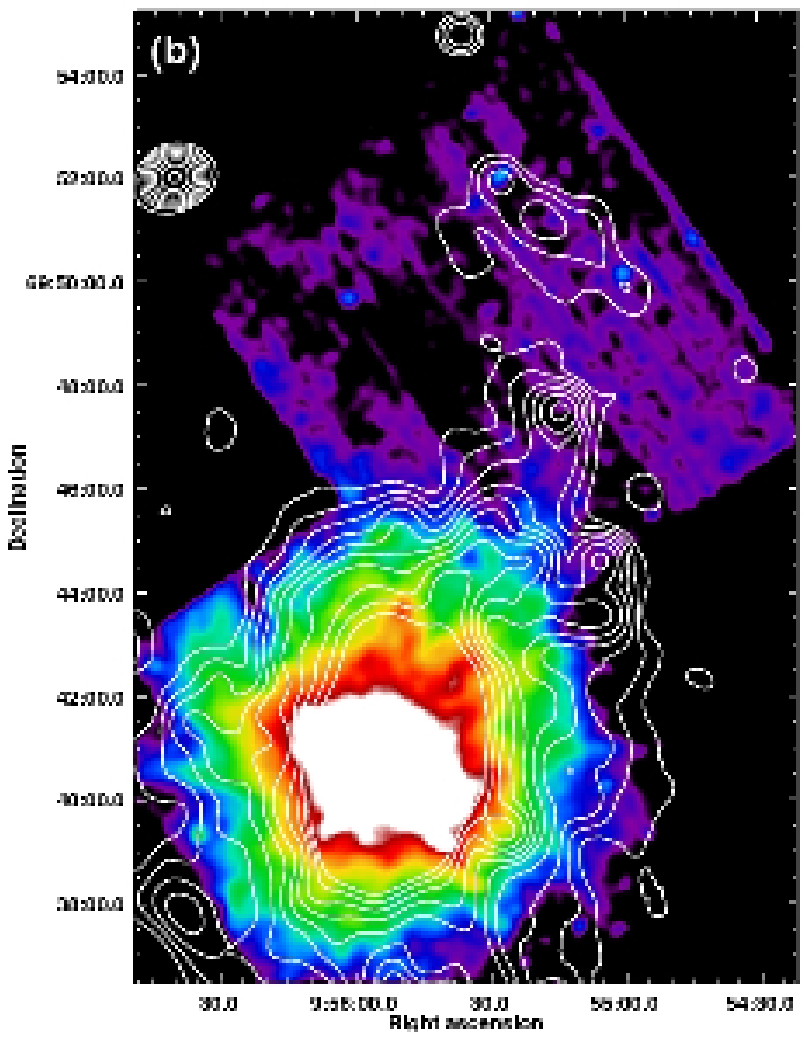}
\includegraphics[width=0.33\textwidth]{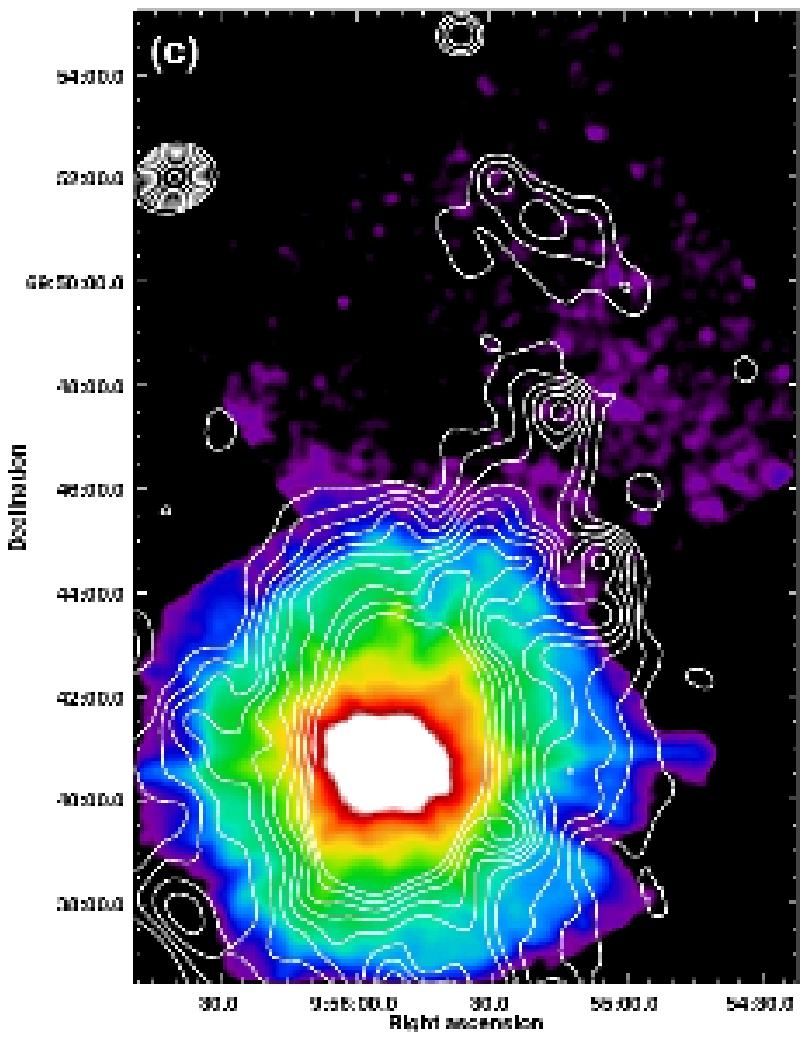}\\
\includegraphics[width=0.33\textwidth]{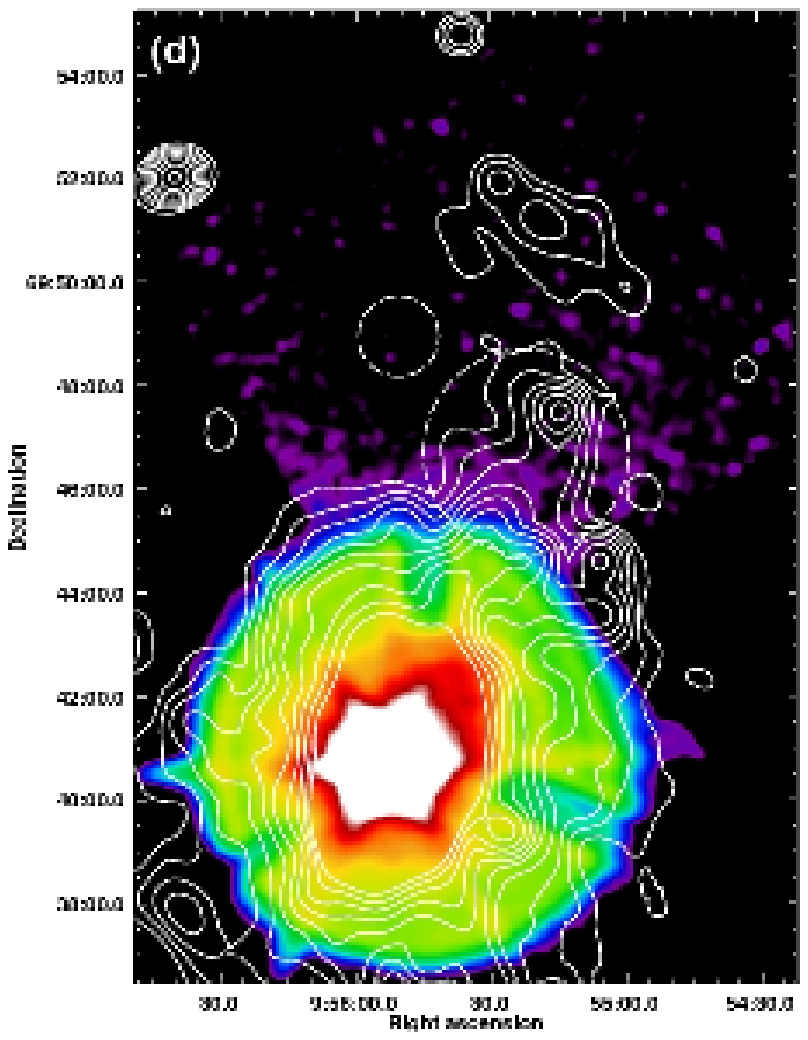}
\includegraphics[width=0.33\textwidth]{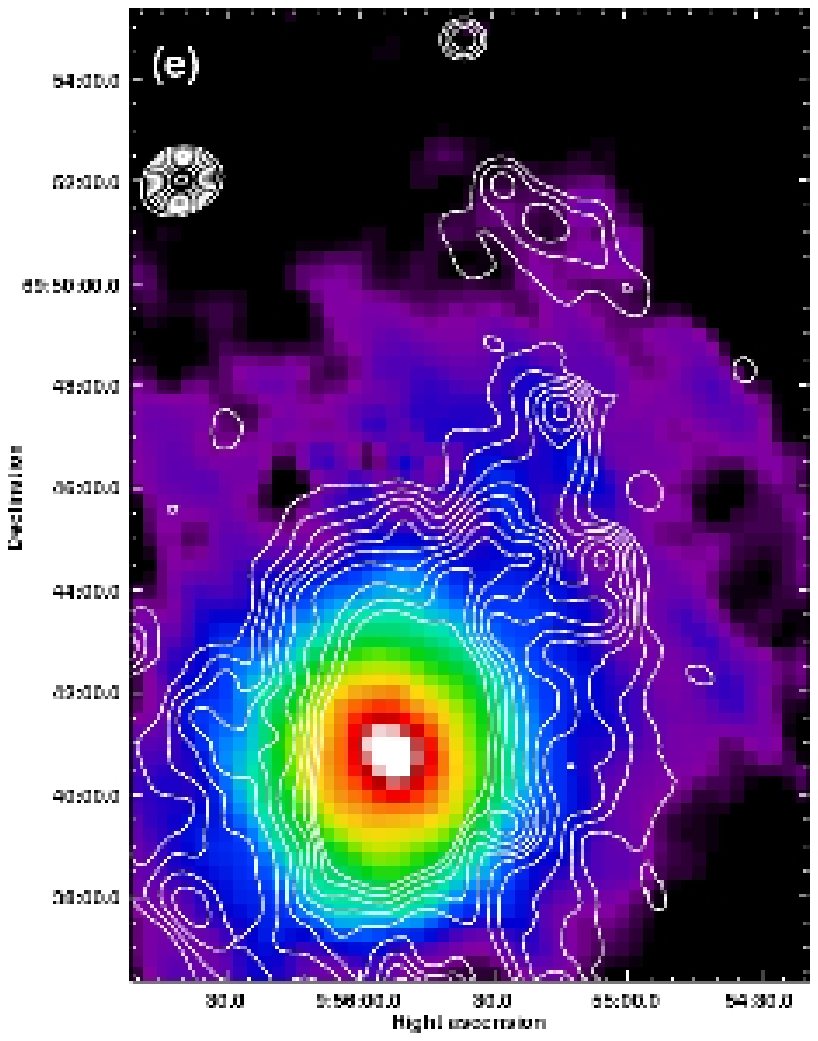}
\includegraphics[width=0.33\textwidth]{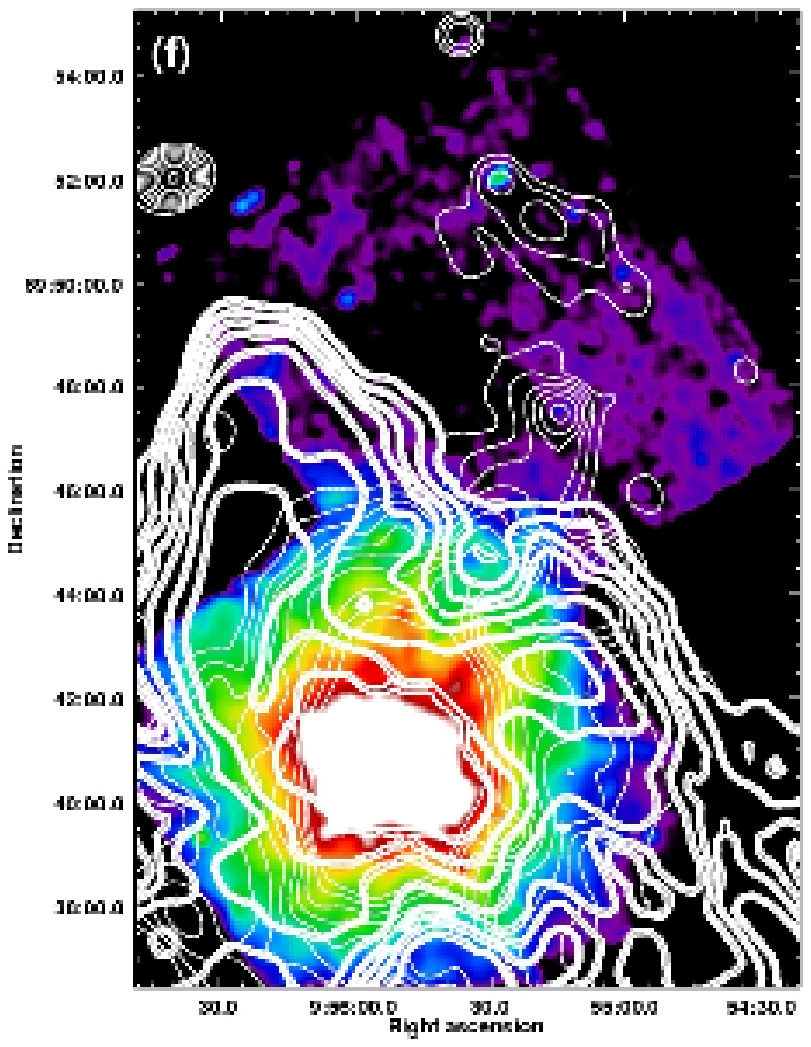}
\caption{Low-level MIR and FIR images of M~82 including the Cap region in the (a) $S7$, (b) $S11$, (c) $L15$, (d) $L24$, (e) $N160$ bands, overlaid on the XMM/Newton X-ray (0.2--10 keV) contour map in a logarithmic scale. (f) The \ion{H}{i} contour map of M~82 in the thick lines, taken from Yun et al. (1994), is superposed on the panel (a) image. The color scales of the MIR and FIR images are logarithmically scaled from 0.2 \% to 0.0007 \% of the peak surface brightness for all the panels except (e) and from 80 \% to 0.1 \% for the panel (e). The two circular apertures are shown in the panel (d), where the larger and the smaller one are used to obtain the flux densities from the halo ($d\leq 4'$) regions and to estimate the background level and its fluctuation from the darkest nearby blank sky, respectively. }
\end{figure*}

\subsection{Spectral energy distributions}
We derive the flux densities of M~82 by integrating the surface brightness within circular apertures of diameters of $2'$ ($\sim$2 kpc), $4'$, and $8'$ around the center of the galaxy in each photometric band image of Figs.1 and 2. For the IRC, no aperture corrections are performed since above aperture sizes are sufficiently large as compared to the PSF \citep{Ona07}. For the FIS, point-source aperture corrections are performed by using the correction table given in Shirahata et al. (2009), where the correction factors include the effects of the optical cross talk in the $N60$ and $WIDE$-$S$ images. We also derive the flux densities of the central regions within a diameter of $1'$ for the IRC and 12\farcm5 for the FIS; the latter is the largest aperture covered by all the 4 FIR bands. For the former, the aperture size is as small as the FWHMs ($37''-61''$) of the PSFs for the FIS (Kawada et al. 2007) and thus we cannot obtain reasonable fluxes in the FIR since the central emission is not a point source.

We obtain the flux densities of the halo region within a diameter of $4'$, which is located at R.A. (J2000) $=$ 9 55 21.0 and Dec. (J2000) $=$ $+$69 46 46.0. The two 4-kpc apertures in the center and the halo are shown on the FIR image in Fig.2a and the same aperture in the halo is also indicated on the MIR image in Fig.3d. The flux densities in all the MIR and FIR bands thus obtained for the center ($d\leq 2'$, $d\leq 4'$), the halo ($d\leq 4'$), and the total region ($d\leq 8'$) are listed in Table 2.
   
The spectral energy distributions (SEDs) constructed from the flux densities in Table 2 are presented in Fig.4. The SEDs of the regions centered at the nucleus with diameters of $1'$, $2'$, $4'$, $8'$, and 12\farcm5 are shown in the ascending order in Fig.4a, while the SED of the center ($d\leq 2'$), the differential SEDs of $4'-2'$, $8'-4'$, and 12\farcm5$-8'$, and the SED of the halo region ($d\leq 4'$) are given in the descending order in Fig.4b, in units of Jy arcmin$^{-2}$ (flux densities divided by the corresponding area). The figures clearly show that the SED is getting softer toward regions farther away from the galactic center. In particular, the lowest two FIR SEDs in Fig.4b, for which the integrated areas are partially overlapped, indicate the presence of cold dust in the halo.

\begin{table*}
\caption{Flux densities of M~82}
\label{flux}
\centering
\renewcommand{\footnoterule}{}
\begin{tabular}{lrrrrrr}
\hline\hline
Band & Center ($d\leq 1'$) & Center ($d\leq 2'$) & Center ($d\leq 4'$)$^{\mathrm{a}}$ & Total ($d\leq 8'$) & Total ($d\leq$ 12\farcm5) &Halo ($d\leq 4'$)$^{\mathrm{a}}$\\
 & (Jy) & (Jy) & (Jy) & (Jy) & (Jy )& (Jy)\\
\hline
IRC $S7$ 7 $\mu$m & 54.1$\pm$1.2$^{\mathrm{b}}$ & 64.2$\pm$1.5 & 72.7$\pm$1.7 & 75.3$\pm$1.7 &\dots & 0.045$\pm$0.001 \\
IRC $S11$ 11 $\mu$m & 43.7$\pm$1.0 & 53.8$\pm$1.3 & 62.8$\pm$1.5 & 65.5$\pm$1.5 &\dots & 0.048$\pm$0.001 \\
IRC $L15$ 15 $\mu$m & 112.0$\pm$3.2 & 128.0$\pm$3.6 & 137.0$\pm$3.9 & 140.8$\pm$4.0 &\dots & 0.091$\pm$0.002 \\
IRC $L24$ 24 $\mu$m & 307$\pm$14 & 342$\pm$16 & 361$\pm$17 & 365$\pm$17 &\dots & 0.21$\pm$0.01 \\
FIS $N60$ 65 $\mu$m &\dots & 630$\pm$130 & 1530$\pm$310 &1700$\pm$340 & 1920$\pm$380 & 16.4$\pm$3.3 \\
FIS $WIDE$-$S$ 90 $\mu$m &\dots  & 529$\pm$110 & 1550$\pm$310 & 1840$\pm$370 & 2040$\pm$410 & 16.0$\pm$3.2 \\
FIS $WIDE$-$L$ 140 $\mu$m &\dots & 520$\pm$160 & 1580$\pm$480 & 2000$\pm$600 & 2250$\pm$670 & 26.2$\pm$7.9 \\
FIS $N160$ 160 $\mu$m &\dots & 360$\pm$110 & 1270$\pm$380 & 1580$\pm$470 & 1940$\pm$580 & 47$\pm$14 \\
\hline
\end{tabular}
\begin{list}{}{}
\item[$^{\mathrm{a}}$] The sizes and positions of the aperture regions are indicated by the dashed circles in Figs.2a and 3d.  
\item[$^{\mathrm{b}}$] The flux density errors include both systematic effects associated with the detectors and absolute calibration uncertainties (Lorente et al. 2007; Verdugo et al. 2007).
\end{list}
\end{table*}

\begin{figure*}
\includegraphics[width=0.5\textwidth]{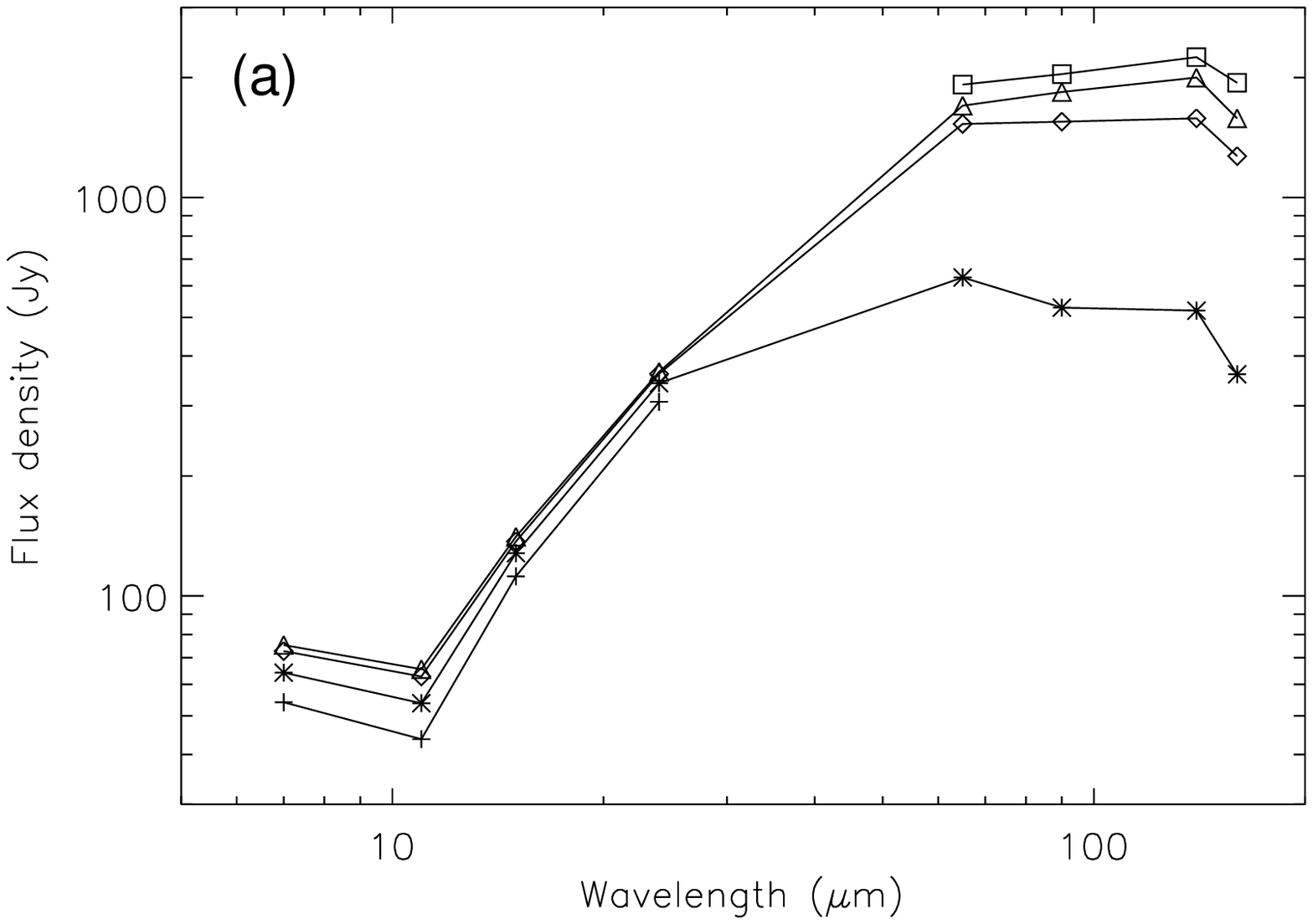}
\includegraphics[width=0.5\textwidth]{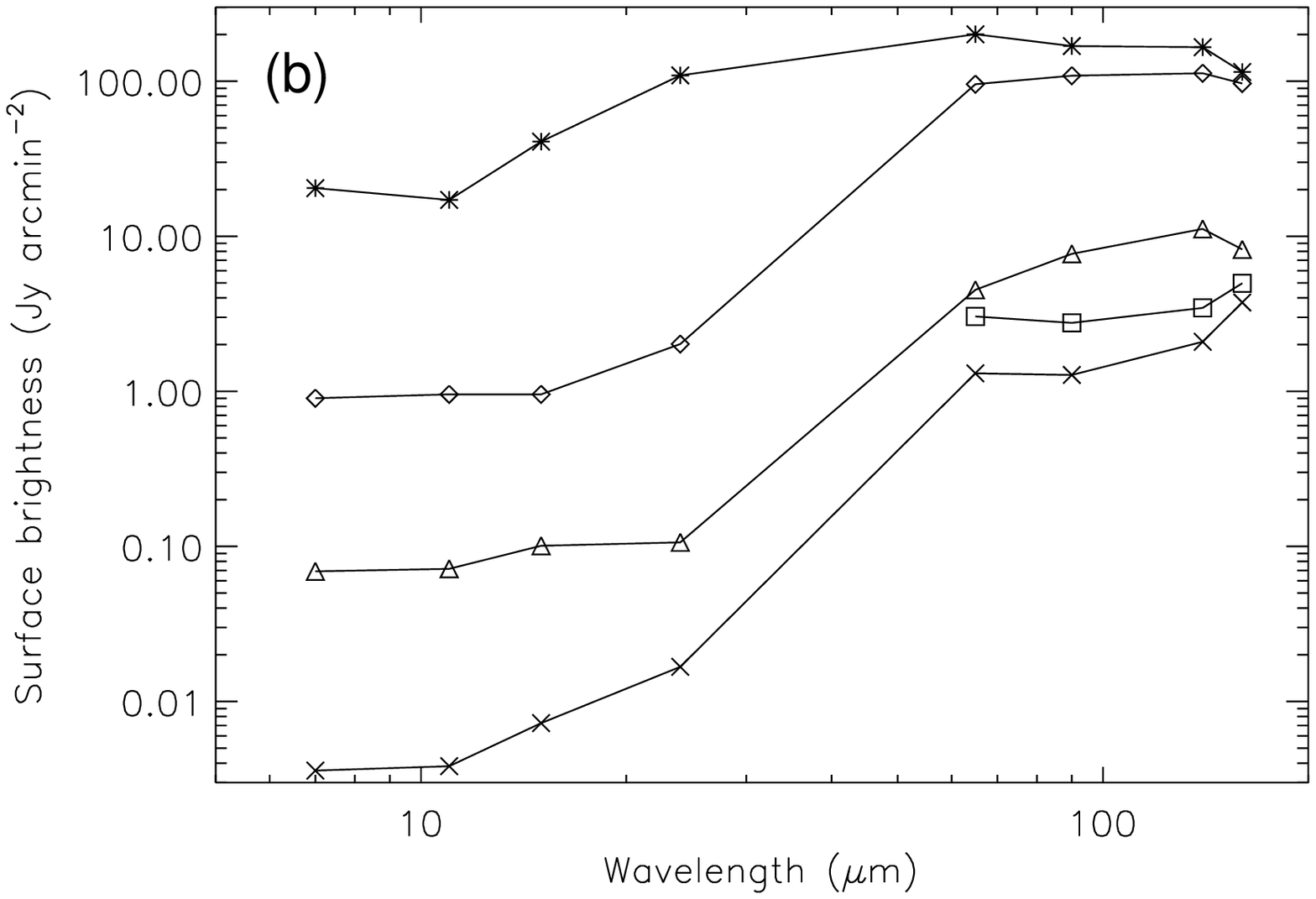}
\caption{Spectral energy distributions (SEDs) created from the flux densities given in Table 2. (a) The SEDs of the regions centered at the nucleus with diameters of $1'$, $2'$, $4'$, $8'$, and 12\farcm5 in the ascending order. (b) The SED of the center ($d\leq 2'$), the differential SEDs of $4'-2'$, $8'-4'$, and 12\farcm5$-8'$, and the SED of the halo region ($d\leq 4'$) in the descending order, given in units of Jy arcmin$^{-2}$. }
\end{figure*}

\subsection{Comparison with other wavelength images}
Figure 5 shows comparison between the $S7$ contour map and the continuum-subtracted H$\alpha$ image; the former is the same as presented in Fig.1a, while the latter is taken from the NASA/IPAC Extragalactic Database. We find that they are remarkably similar to each other. The H$\alpha$ image is dominated by a biconical structure that defines the superwinds. The excellent spatial correspondence indicates that the PAHs responsible for the $S7$ band flux are well mixed with the ionized gas and entrained by the galactic superwinds. Engelbracht et al. (2006) pointed out that the morphology of the Spitzer 8 $\mu$m image of M~82 is similar to that of the H$\alpha$ emission, but also that the 8 $\mu$m image differs in that the emission is bright all around the galaxy rather than being dominated by a cone perpendicular to the disk. As far as the central $7'$ ($\simeq$ 7 kpc) area is concerned, however, our result reveals a marked similarity and no significant deviation in morphology between the PAH and the H$\alpha$ emission. 

We calculate the linear-correlation coefficient, $R$, between the low-level $S7$ and H$\alpha$ images for a bin size of 2\farcs3 and the $S7$ brightness ranging from 1 \% to 0.01 \% of the peak. We obtain $R=+0.80$ (for a total of 21014 data points). In general, as observed in our Galaxy and nearby galaxies, most PAHs are associated with neutral gas and their emission is very weak in ionized regions with strong radiation field probably due to destruction \citep[e.g.][]{Bou88,Des90,Ben08}. Therefore the strong positive correlation observed between the PAH and H$\alpha$ emission is rather unusual. This may be due to a difference in the length of time over which PAHs have been exposed to a harsh environment. The tight correlation would then suggest that the PAHs have been traveling fast enough to reach their present locations, $\sim 3$ kpc above the disk, in less than a destruction timescale.

Figure 6 shows comparison between the $WIDE$-$L$ and the \ion{H}{i} contour map; the former is the same but enlarged $25'\times 25'$ image as presented in Fig.2c, while the latter is taken from Yun et al. (1994). The \ion{H}{i} map shows the large-scale streamers expanding almost in parallel to the disk. In addition, there are at least three prominent extending \ion{H}{i} structures toward the north, west, and southeast directions in spatial scales of $5'-10'$, which might be related to the three filamentary structures in the PAH emission. Although the overall morphology of the FIR image is somewhat different from the \ion{H}{i} map, the FIR-bright regions at $12'$ to the west and $8'$ to the southeast as well as the extended structures connecting to them from the center show some spatial resemblance to the \ion{H}{i} distribution. The difference in overall distribution between the dust and the \ion{H}{i} gas might be explained by inhomogeneity of the intergalactic UV radiation heating the dust, which is likely attenuated more along the major axis than the minor axis. The large-scale neutral streamers are probably caused by a past tidal interaction of M~82 with M~81 (Yun et al. 1993). The intergalactic FIR dust can thus be attributed to leftover clouds ejected out of the galaxy by the tidal interaction and residing in the intergalactic medium.

In Fig.3, we superpose an X-ray contour map of M~82 from the XMM/Newton archival data on the MIR and FIR images. The conspicuous X-ray superwind extends along the northwest direction, correlating very well with the FIR emission. We obtain $R=+0.60$ ($N=2417$) between the low-level FIR ($N160$) and X-ray images for a bin size of $15''$, specifically the $N160$ brightness higher than 0.1 \% (the lowest color level in Fig.3e), and the X-ray brightness lower than 0.1 \% of the peak (an intermediate contour level in Fig.3e). The spatial correlation between X-ray and FIR suggests that the FIR dust is entrained by the superwind and outflowing from the galactic plane. 

In contrast, there is no clear correlation between the X-ray superwind and the PAH/VSG emission. For example, we obtain $R=+0.21$ ($N=13568$) between the low-level $S7$ and X-ray images for a bin size of $4''$, the $S7$ brightness higher than 0.0007 \% (the lowest color level in Fig.3a), and the X-ray brightness lower than 0.1 \% of the peak (the same as the above). There might be even some anti-correlation between them especially in the $S7$ image; the X-ray plasma seems to be situated in between the two filamentary structures in the PAH emission. Moreover in the Cap region, where there is significant diffuse MIR emission, the PAH emission is somewhat reduced locally at the position of the X-ray Cap (see section 4.3). This might reflect that the PAHs in preexisting diffuse neutral clouds are destroyed by collision of the energetic superwind.   
The difference between the PAHs and the FIR dust indicates that the PAHs are more easily destroyed in hot plasma.

\begin{figure}
\includegraphics[width=0.5\textwidth]{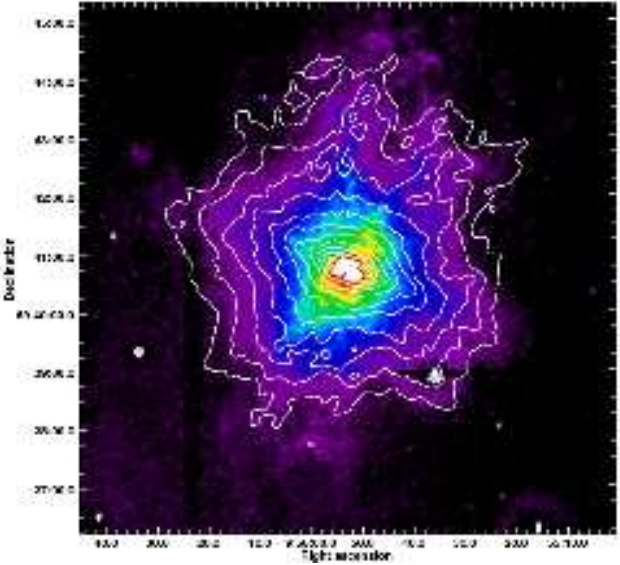}
\caption{Continuum-subtracted H$\alpha$ image of M~82 superposed on the same $S7$ contour map as Fig.1a. }
\end{figure}

\begin{figure}
\includegraphics[width=0.5\textwidth]{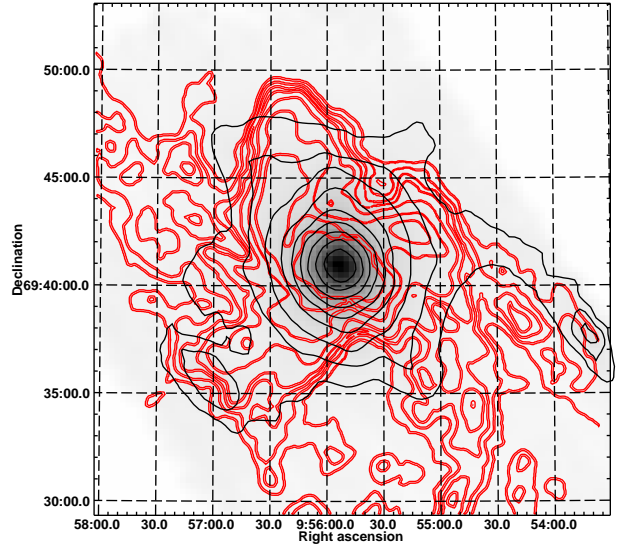}
\caption{\ion{H}{i} contour map of M~82 taken from Yun et al. (1994) in the thick red lines, the same as in Fig.3f, superposed on the same $WIDE$-$L$ contour map as Fig.2c. }
\end{figure}

\section{Discussion}
\subsection{PAH properties}
 The striking similarity between the $S7$ and $S11$ band images in Fig.1 indicates that spatial variations in properties of PAHs are quite small; the PAH interband strength ratio of the $6-9$ $\mu$m to the $11-13$ $\mu$m emission is nearly constant throughout the observed regions of M~82. In Fig.7, we calculate the ratios of surface brightness at 11 $\mu$m to that at 7 $\mu$m. Prior to dividing the images, we have applied smoothing to both images by a gaussian kernel of $24''$ in FWHM. The area in the $S7$ image with brightness levels lower than 1.0 MJy/str that corresponds to the lowest contour in Fig.1 is masked in calculating the ratios. As a result, the ratio map in Fig.7 reveals that the values are nearly constant at $\sim$ 0.8, which is consistent with the spectroscopic results on M~82 by ISO \citep{For03b} and Spitzer \citep{Bei08}.

The relative strength of the different PAH bands is expected to vary with the size and the ionization state of the PAHs \citep[e.g.][]{Dra07}. The C-C stretching modes at 6.2 and 7.7 $\mu$m are predominantly emitted by PAH cations, while the C-H out-of-plane mode at 11.3 $\mu$m arises mainly from neutral PAHs \citep{All89,Dra07}. Neutral PAHs emit significantly less in the 6--8 $\mu$m emission \citep{Job94,Kan08a}. The size effect on the PAH 11.3 $\mu$m/7.7 $\mu$m ratio is relatively small \citep{Dra07}, although, in general, smaller PAHs can emit features at shorter wavelengths. 
Hence, the variations of the $S11$/$S7$ ratio in the halo regions where there is no extinction effects show support for variation mostly in PAH ionization, i.e. a larger fraction of neutral PAHs for higher $S11$/$S7$ ratios, with small contributions from changes in PAH size distribution. As seen in Fig.7, regions with higher ratios have a reasonable tendency to be located in the outskirts of the PAH emission distribution, where the UV radiation is weaker. Apart from the overall tendency, there is a region showing systematically higher $S11$/$S7$ ratios toward the east direction from the center, where the UV radiation field is expected to be relatively weak in such a large scale. This relative weakness might reflect the spatial distribution of dense gas near the central region shielding UV light from the nucleus; the central 3 kpc CO map of M~82 in Walter et al. (2002) exhibits a distribution of molecular gas with larger viewing angles toward the east direction from the nucleus. 

Beir\~ao et al. (2008) observed an enhancement of the 11.3 $\mu$m PAH feature relative to the underlying continuum emission outward from the galactic plane, and suggested that the UV radiation field excites PAHs and VSGs differently. In the halo regions, the radiation is still intense enough to excite PAHs but no longer intense enough to excite VSGs to the same temperatures as in the galactic plane. Thus it is reasonable that the $S7$ and $S11$ images show more extended emission than the $L15$ and $L24$ images in Fig.1. Moreover the ratio map in Fig.7 suggests that the radiation field is a little weaker in the northeast filament than in the northwest filament, which may explain the difference in the PAH and VSG emission intensity between these filaments (Fig.1).

\begin{figure}
\includegraphics[width=0.5\textwidth]{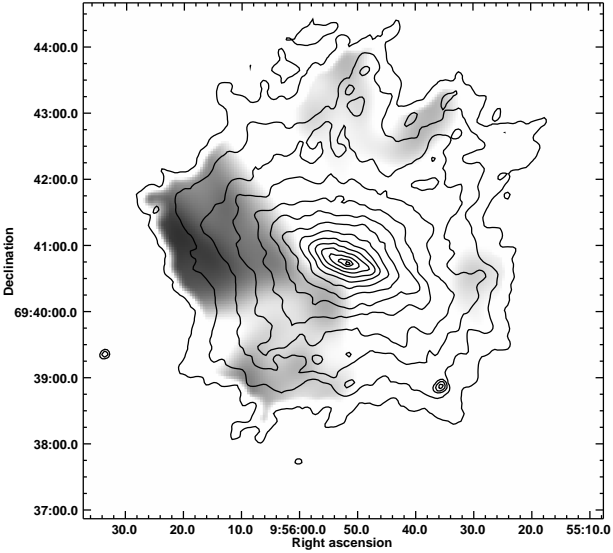}
\caption{Ratios of the $S11$ to the $S7$ band surface brightness in grey scales linearly drawn from 0.9 (white) to 1.3 (black), overlaid on the same $S7$ contour map as Fig.1a. }
\end{figure}

\subsection{Dust mass and luminosity}
We fit the SEDs in Fig.4 by a three-temperature dust grey-body plus PAH component model (Fig.8). The PAH parameters are taken from Draine \& Li (2007) by adopting the PAH size distribution and fractional ionization typical of diffuse ISM and assuming the interstellar radiation field in the solar neighborhood. The latter assumption does not influence the spectral shape much unless the radiation field is as much as $10^4$ times higher \citep{Dra07}, while Colbert et al. (1999) estimated $2.8$ times higher than the solar neighborhood value from photodissociation regions in the central $\sim 1$ kpc area of M~82.

For the dust grey-body model, we adopt an emissivity power-law index of $\beta=1$ for every component. We started with the initial conditions of 140 K, 60 K, and 28 K for the dust temperatures of the three components. These components are just representatives to reproduce the shape of the dust continuum spectra as precisely as possible and we below consider only the total luminosity and dust mass by summing up the corresponding values from the three components. We calculate dust mass by using the equation \citep[e.g.][]{Hil83}:
\begin{equation}
M_{\rm dust}=\frac{4a\rho D^2}{3}\frac{F_{\nu}}{Q_{\nu}B_{\nu}(T)},
\end{equation}
where $M_{\rm dust}$, $D$, $a$, and $\rho$ are the dust mass, the galaxy distance, the average grain radius, and the specific dust mass density, respectively. $F_{\nu}$, $Q_{\nu}$, and $B_{\nu}(T)$ are the observed flux density, the grain emissivity, and the value of the Planck function at the frequency of $\nu$ and the dust temperature of $T$. We adopt the grain emissivity factor given by Hildebrand (1983), the average grain radius of 0.1 $\mu$m, and the specific dust mass density of 3 g cm$^{-3}$. We use the 90 $\mu$m flux density and the dust temperature, both given by each dust component. The dust mass thus derived as well as the total luminosity of the PAH and dust components, $L_{\rm PAH}$ and $L_{\rm dust}$, are listed in Table 3. We should note that the spectral fitting to the SED of the total ($d\leq 8$ kpc) region does not require additional component, but the differential SED in Fig.4b shows the presence of colder dust emission peaking at a wavelength longer than 160 $\mu$m. Hence the dust mass in the total region is considered to be lower limits to the real dust mass contained in this region.

The ISO/LWS spectroscopy of the central $\sim 1$ kpc region of M~82 showed that the SED over the wavelength range of 43-197 $\mu$m is well fitted with a 48 K dust temperature and $\beta=1$, giving a total infrared flux of $3.8\times 10^{10}$ $L_{\odot}$ \citep{Col99}. Our result shows overall consistency with the ISO result. Thuma et al. (2000) estimated from their 240 GHz measurement that the total dust mass in the inner 3 kpc of the galaxy is $7.5\times 10^6$ $M_{\odot}$, which is in an excellent agreement with our result. 
The total mass of atomic hydrogen gas in M~82 is estimated to be $8\times 10^8$ $M_{\odot}$ \citep{Yun94}, while the mass of molecular gas in an area of $2.8\times 3.9$ kpc$^2$ is $1.3\times 10^9$ $M_{\odot}$ \citep{Wal02}. Hence the gas-to-dust mass ratio for the total 8 kpc region is $\sim$ 200, similar to the accepted value of $100-200$ for our Galaxy \citep{Sod97}. In the inner $2$ kpc, the molecular gas mass is estimated to be $1.0\times 10^9$ $M_{\odot}$ from Fig.1 and Table 1 in Walter et al. (2002), and thus the gas-to-dust mass ratio is $\sim$400, where contribution of atomic hydrogen gas is not included. Hence the gas-to-dust ratio is considerably higher in the central region than in the halo; similar results were also obtained for NGC~253 \citep{Rad01,Kan09b}. The possible presence of colder dust in the halo would further increase the difference. In much larger spatial scales, Xilouris et al. (2006) obtained the \ion{H}{i}-to-dust mass ratio of $\sim$20 from a systematic shift in the color of background galaxies viewed through the intergalactic medium of the M~81-M~82 group.

The variations of $L_{\rm PAH}$, $L_{\rm dust}$, and $M_{\rm dust}$ in Table 3 reveal a rapid decrease in the radiation field intensity but not in the dust mass away from the galactic plane. The ratio $L_{\rm PAH}$/$L_{\rm dust}$ is fairly constant ($\sim 0.1$) for all the regions except the central 2 kpc region where $L_{\rm PAH}$/$L_{\rm dust}$ is about 0.2; the PAH emission in the central region is relatively bright due to the nuclear starburts. The relatively small change in $M_{\rm dust}$ indicates that the total amount of the dust residing in the halo is so large that it can be comparable to that of the dust contained in the galaxy body. Hence the dust-enriched gas injected by the starburst significantly contributes to the total mass of wind material that is tranported out of the disk.

\begin{figure*}
\includegraphics[width=0.5\textwidth]{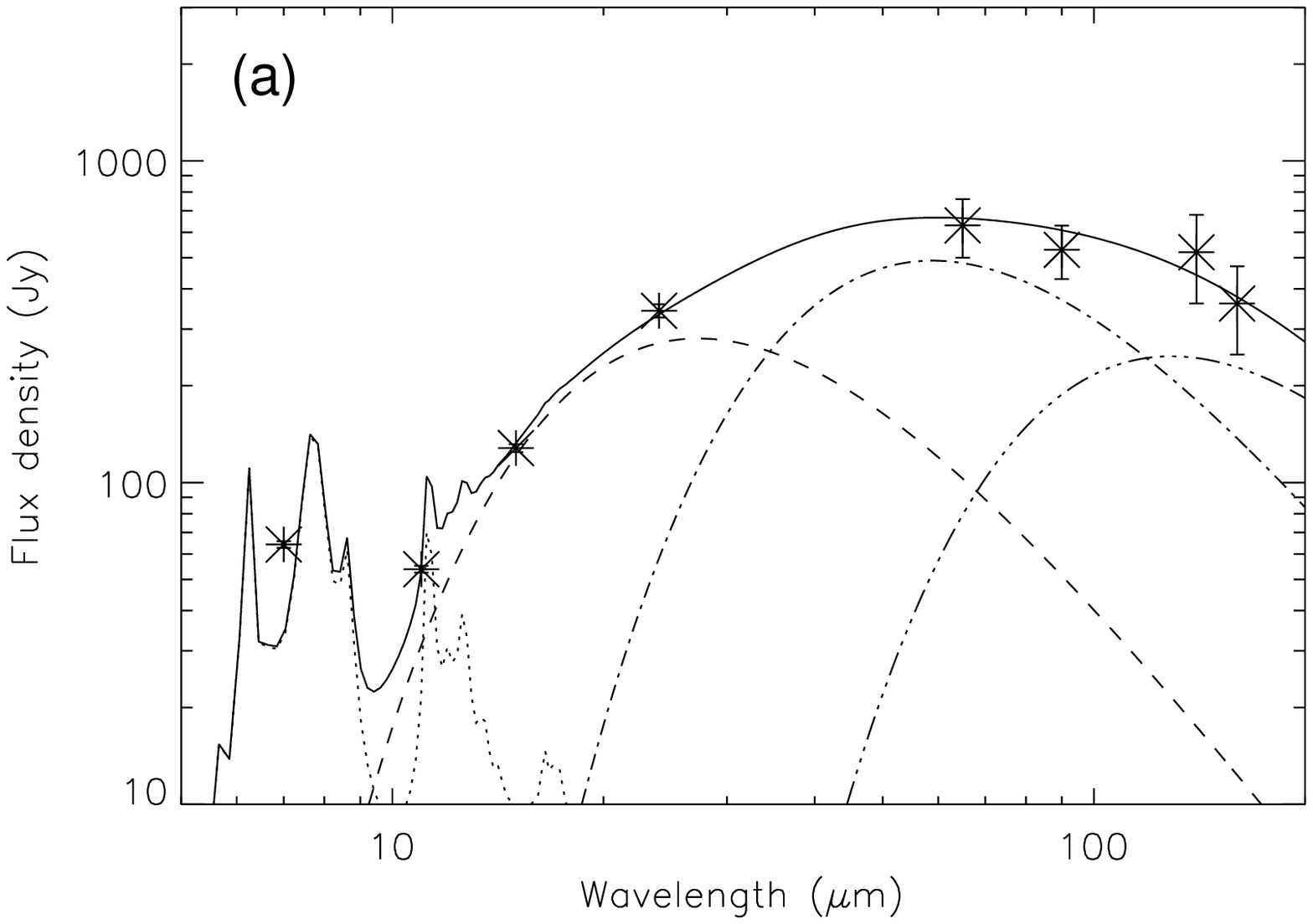}
\includegraphics[width=0.5\textwidth]{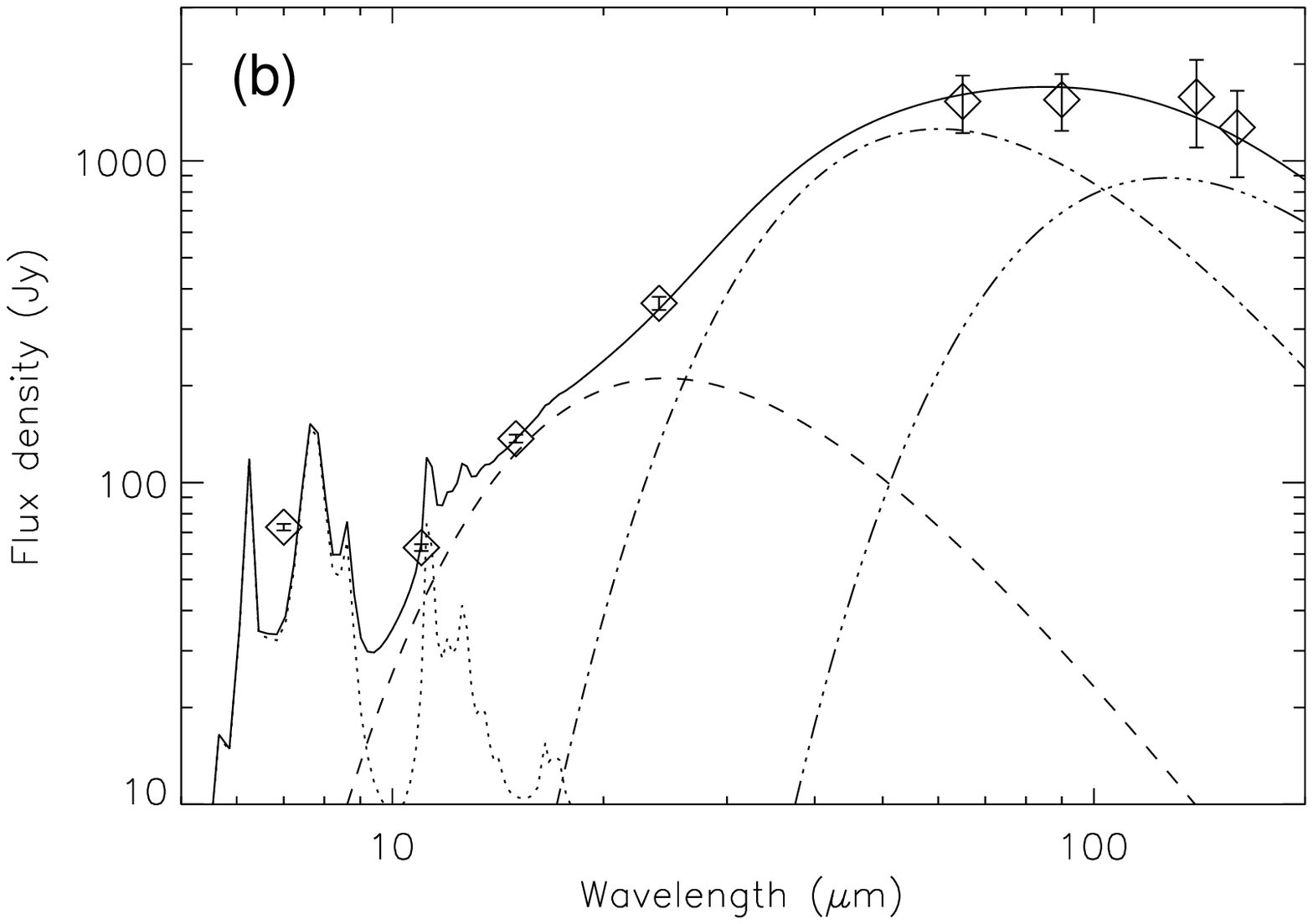}\\
\includegraphics[width=0.5\textwidth]{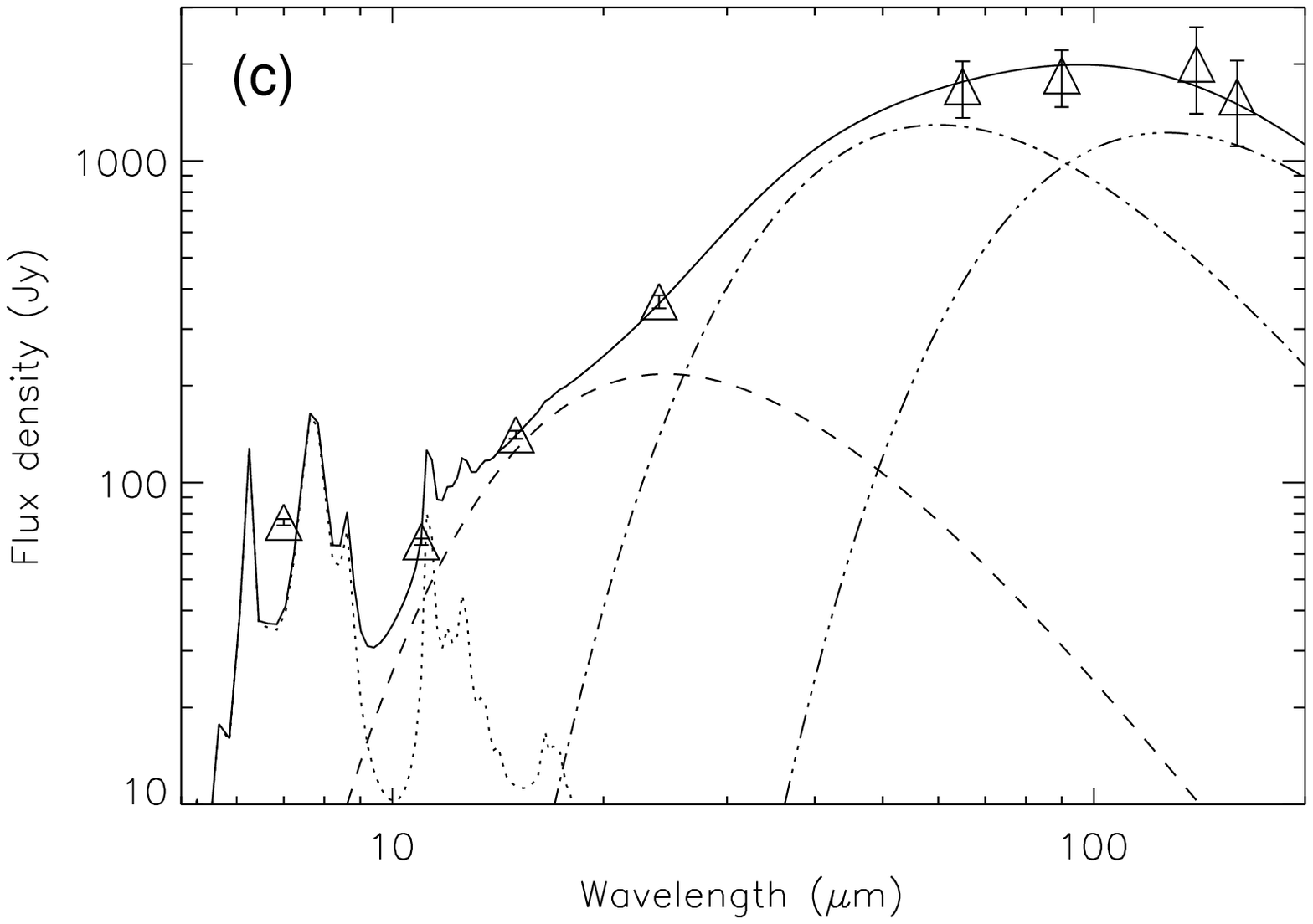}
\includegraphics[width=0.5\textwidth]{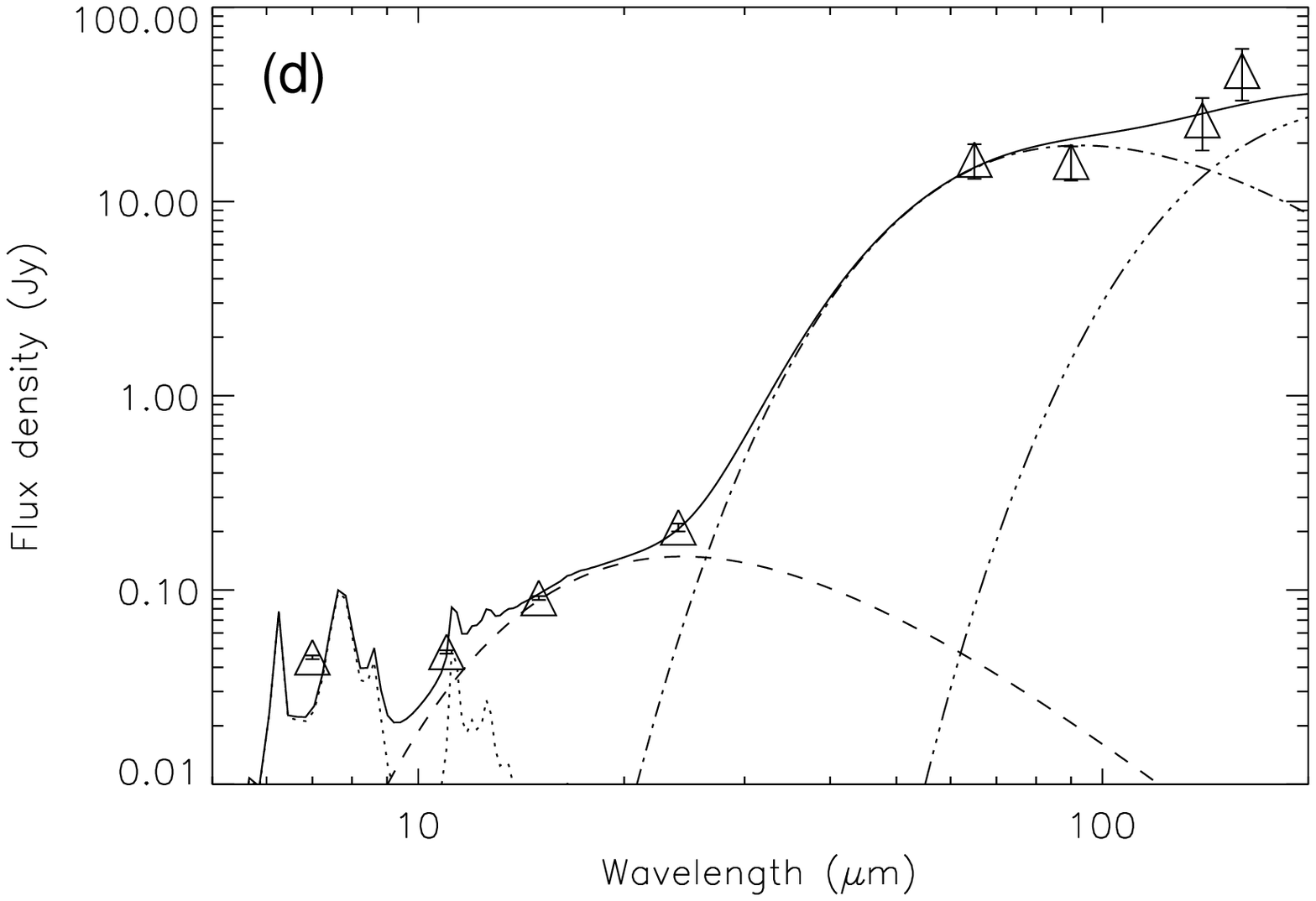}
\caption{Fitting results by a three-temperature dust grey-body plus PAH component model to the SEDs in Fig.4 for the (a) 2 kpc, (b) 4 kpc, and (c) 8 kpc central regions as well as the (d) 4 kpc halo region. }
\end{figure*}

\subsection{Interplay of dust and PAHs with various gas components}
There are at least three filamentary structures of the PAH emission (Fig.1).
%, among which only the southern filament is detected in the submillimeter and CO emission (Leeuw \& Robson 2009; Thuma et al. 2000). 
The tight correlation between the PAH and H$\alpha$ emission provides evidence that the PAHs are well mixed in the ionized superwind gas and outflowing from the disk. In contrast, the H$_{2}$ 2.12 $\mu$m emission \citep{Vei09} shows a relatively loose correlation with the PAH emission in the northern halo, while they are well correlated in the southern halo. Veilleux et al. (2009) suggested that UV radiation is important for the excitation of the warm H$_2$ in the southeast extension because this is a region where photoionization by OB stars dominates over shocks \citep{Sho98}, whereas a dominant heating mechanism is not well determined among shock, UV pumping, and X-ray heating in the northern halo. Our result favors the scenario that shock or X-ray heating is more important for the warm H$_2$ in the northen halo rather than the UV radiation that heats the PAHs, which could explain the loose correlation between the H$_2$ and PAH emission in the northern halo.  

The deprojected outflow velocity of the H$\alpha$ filaments is $525-655$ km s$^{-1}$ \citep{Sho98}. Therefore the PAHs seem to have survived in a harsh environment for about 5 Myr to reach the observed positions at $\sim 3$ kpc above the plane.   
The dominant excitation mechanism for the H$\alpha$ filaments is most likely photoionization by the nuclear starburst; UV radiation is escaping from the disk along a channel excavated by the hot superwind \citep{Sho98}. Shock ionization begins to contribute toward larger radii, and beyond 4 kpc above the disk in the northern halo, X-ray hot plasma is a dominant phase of the superwind. In the hot plasma phase, the PAH emission is significantly reduced, as can be seen in Fig.3,  which might imply that the PAHs are destroyed in the hot plasma. This would happen in $\sim$6 Myr after being ejected with a hot plasma whose outflow velocity is $\sim$ 700 km s$^{-1}$ \citep{Leh99}. 

Consistently with the fact that the dust temperature is expected to decrease with attenuating radiation field away from the disk, the VSG emission cannot be seen in the halo beyond 4 kpc above the disk, either (Fig.3). Nevertheless, as seen in Fig.3e, the FIR dust emission is still observed in the hot plasma superwind beyond 4 kpc. As discussed in Tsuru et al. (2007), the lifetime of dust of a 0.1 $\mu$m size against sputtering destruction in the hot plasma superwinds of M~82 is 20$\times f^{0.5}$ Myr, where $f$ is the volume filling factor of the hot plasma that is expected to be as low as $\sim$ 0.01 \citep{Str00}. If we assume that the observed dust is homogeneously mixed in the hot plasma, the dust grains have to survive during a traveling time of $5-10$ Myr to reach their present locations, which is comparable to the sputtering destruction time scale. Indeed we have direct evidence that there is a close morphological correpondence between the PAHs/dust and the hotter phases of the galactic winds probed in the H$\alpha$ and X-ray emission. 
%Due to poor spatial resolution in the FIR, however, there is no information on how similar in distribution the FIR dust emission is to the above PAH and $H\alpha$ filaments. 

As seen in Fig.6, the extended FIR dust emission is also detected from part of the \ion{H}{i} streamers, the surface of which is probably illuminated by UV light escaping from the disk along a channel excavated by the hot superwind. The streamers are thought to provide evidence that the gas within the optical disk of M~82 is disrupted by the interaction of M~82 with M~81 100 Myr ago, and likely triggers the starburst activitiy in the center of M~82 \citep{Wal02}. Furthermore, the PAH emission in the halo also exhibits significant enhancement in surface brightness in the eastern edge of the Cap region, which spatially corresponds to the edge of the \ion{H}{i} clouds (Fig.3f). Hence the \ion{H}{i} streamers should contain dust and PAHs, obviously not of a primordial origin, but rather as leftover clouds of the past interaction of M~82 with M~81. Detailed modeling of starburst activity in M~82 on the basis of NIR-MIR spectroscopy suggested the occurrence of starburst in two successive episodes, about 10 and 5 Myr ago, each lasting a few million years \citep{For03a}. From the Spitzer/IRS spectroscopy, Beir\~ao et al. (2008) indicated that the star formation rate has decreased significantly in the last 5 Myr. The above dynamical time scales of the superwinds are consistent with these results. The inclusion of the dust and PAHs in the \ion{H}{i} streamers implies that the streamers were already enriched by metals prior to the starburst episodes at the M~82 nucleus. 

As for the Cap region, we detect significant signals at wavelengths of 7 and 11 $\mu$m, which are diffusely extended near the X-ray and H$\alpha$ Cap but appear to be reduced locally at the position of the Cap (Figs.3a and 3b). Lehnert et al. (1999) suggested that the Cap is the result of a collision between the hot superwind and a preexisting neutral cloud. We adopt a size of $3.7\times 0.9$ kpc$^2$ for the X-ray Cap according to Lehnert et al. (1999). By integrating signal within the $3.7\times 0.9$ kpc$^2$ aperture centered at R.A. (J2000) $=$ 9 55 17.0 and DEC (J2000) $=$ $+$69 51 13.0, the X-ray peak position of the Cap, we obtain the flux densities of 4.4 mJy and 9.1 mJy at 7 $\mu$m and 11 $\mu$m, respectively. From the same aperture but at the different position of R.A. (J2000) $=$ 9 54 48.0 and DEC (J2000) $=$ $+$69 48 26.0 that is located in the middle of the faint diffuse emission region, we obtain the flux densities of 8.8 mJy and 11 mJy at 7 $\mu$m and 11 $\mu$m, respectively. The 1-sigma statistical errors are estimated to be 0.3 mJy and 0.5 mJy at 7 $\mu$m and 11 $\mu$m, respectively, from the nearby darkest blank sky. Therefore the signal reduction at the Cap seems to have a statistical significance, which is higher at 7 $\mu$m possibly reflecting that PAHs of smaller sizes are easier to be destroyed there. By adopting the flux ratio between the $S7$ and the $N160$ band in the annular region of radii $2'$ to $4'$ from the galactic center, we estimate the $N160$ surface brightness to be $\sim$2.6 MJy sr$^{-1}$ at the position of the Cap, which is only 1.8 times higher than the 1-sigma background fluctuation level in the $N160$ band, and thus consistent with non-detection of FIR signals from the corresponding area.

An integration of the surface brightness of 2.6 MJy leads to the flux density of 0.7 Jy in the $N160$ band for the $3.7\times 0.9$ kpc$^2$ area of the X-ray Cap. By assuming the same FIR dust SED as observed in the 4 kpc halo region (Fig.8d), dust mass in the Cap is estimated to be $6\times 10^4$ $M_{\odot}$. From the $\sim$50 \% signal reduction of the PAH emission at the Cap, a comparable amount of dust might have been already lost there by sputtering destruction. Thus the dust sputtering provides a potential impact on the metal abundances measured for the X-ray plasma in the Cap as pointed out by Tsuru et al. (2007); the masses of Si and Fe in the hot plasma phase are $1.4\times 10^3\times f^{0.5}$ $M_{\odot}$ and $1.8\times 10^3\times f^{0.5}$ $M_{\odot}$, respectively \citep{Tsu07}. The charge-exchange process could be important in X-ray emission from the Cap \citep{Lal04}, where the ionized superwind from M~82 can be assumed to collide with cool ambient gas located at the Cap. From the destruction of the PAHs at the Cap, we expect that the hot plasma is somewhat enriched with carbon there, which might be related to the marginal detection of the \ion{C}{vi} emission line at 0.459 keV due to the charge-exchange process by Tsuru et al. (2007).

We find that observable amounts of dust and PAHs are included in the phases of both ionized (superwinds) and neutral (streamers) gas, which are spatially separated from each other in the northern halo (Fig.3f). A significant fraction of PAHs seem to have been destroyed in the hot plasma phase of the northern superwind beyond 4 kpc from the disk, but still partly remaining in the X-ray Cap. Moreover PAHs seem to be present widely around the Cap region far beyond the disk, which may have been strewn into the intergalactic space by a past tidal interaction with M~81 before the starburst began at the nucleus of M~82.
Hoopes et al. (2005) concluded that the most likely mechanism for the UV emission in the halo of M~82 is scattering of stellar continuum from the starburst by dust in the halo because the brightness of the UV wind is too high to be explained by photoionized or shock-heated gas. Our result supports their conclusion as far as the galactic superwind regions are concerned. But for the Cap at 11 kpc north of M~82, where UV light is also seen in spatial scales similar to the X-ray \citep{Hoo05}, light scattered by dust may not be a substantial component of the UV emission. The diffuse distribution of the intergalactic material represented by PAHs would scatter UV light from the starburst in much wider area rather than the observed region limited to the X-ray Cap. 

\begin{table}
\caption{Luminosities and dust masses derived from the spectral fitting to the observed SEDs}
\label{log}
\centering
\renewcommand{\footnoterule}{}
\begin{tabular}{cccc}
\hline\hline
Region & $L_{\rm PAH}$ & $L_{\rm dust}$ & $M_{\rm dust}$ \\
 & $10^{9}$ $L_{\odot}$ & $10^{10}$ $L_{\odot}$ & $10^{6}$ $M_{\odot}$ \\
\hline
Center ($d\leq 2'$) & 6.0 & 3.1 & 2.3 \\
Center ($d\leq 4'$) & 6.4 & 5.6 & 7.6 \\
Total ($d\leq 8'$)  & 6.9 & 6.1 & 10.3 \\
Halo ($d\leq 4'$) & 0.0041 & 0.0057 & 3.9 \\
\hline
\end{tabular}
\end{table}

\section{Summary}
We have presented new MIR and FIR images of M~82 obtained by AKARI, which reveal both faint extended emission in the halo and very bright emission in the center with signal dynamic ranges as large as five and three orders of magnitude for the MIR and FIR, respectively. Our observations cover wider areas than previous IR observations up to the X-ray/H$\alpha$ Cap at 11 kpc above the disk, which complements previous studies. We detect MIR and FIR emission in the regions far away from the disk of the galaxy, reflecting the presence of dust and PAHs in the halo of M~82. We show that the ionization state of the PAHs is fairly constant throughout the halo of M~82 with small variations in some areas probably due to reduction in the UV radiation escaping from the disk. We find that the dust and PAHs are contained in both ionized and neutral gas components, implying that they have been expelled into the halo of M~82 by both starbursts and galaxy interaction. In particular, we obtain an tight correlation between the PAH and the H$\alpha$ filamentary structures, which provides evidence that the PAHs are well mixed in the ionized superwind gas and outflowing from the disk in a short timescale. We also find that the dust is contained even in the X-ray hot plasma while PAHs are widely spread over the Cap region. Both suggest that the gas in the halo of M~82 is highly enriched with dust, connecting to the results of Xilouris et al. (2006) that dust exists in the intergalactic medium on much larger scales.

\begin{acknowledgements}
We thank all the members of the AKARI projects, particularly those belonging to the working group for the ISMGN mission program. We would also express many thanks to the anonymous referee for giving us useful comments. AKARI is a JAXA project with the participation of ESA. This research is supported by the Grants-in-Aid for the scientific research No. 19740114 and the Nagoya University Global COE Program, "Quest for Fundamental Principles in the Universe: from Particles to the Solar System and the Cosmos", both from the Ministry of Education, Culture, Sports, Science and Technology of Japan.
\end{acknowledgements}


\begin{thebibliography}{}
%\bibitem[Ishihara et al. 2003]{Ishihara03}
%;Ishihara, D., Wada, T., Watarai, H., et al. 2003, \procspie, 4850, 1008
\bibitem[Allamandola et al.(1989)]{All89}
     Allamandola, L. J., Tielens, A. G. G. M., \& Barker, J. R. 1989, \apjs, 71, 733
\bibitem[Alonso-Herrero et al.(2003)]{Alo03}
     Alonso-Herrero, A., Rieke, G. H., Rieke, M. J., \& Kelly, D. M. 2003, \apj, 125, 1210
\bibitem[Alton et al.(1999)]{Alt99}
     Alton, P. B., Davies, J. I., \& Bianchi, S. 1999, \aap, 343, 51
\bibitem[Beir\~ao et al.(2008)]{Bei08}
Beir\~ao, P., Brandl, B. R., Appleton, P. N., et al. 2008, \apj, 676, 304
\bibitem[Bendo et al.(2008)]{Ben08}
     Bendo, G. J., Draine, B. T., Engelbracht, C. W., et al. 2008, MNRAS, 389, 629
\bibitem[Bland \& Tully(1988)]{Bla88}
Bland, J. \& Tully, B. 1988, \nat, 334, 43B
\bibitem[Boulanger et al.(1988)]{Bou88}
     Boulanger, F., Beichman, C., D\'esert, F. X., Helou, G., P\'erault, M., \& Ryter, C. 1988, \apj, 332, 328
\bibitem[Bregman et al.(1995)]{Bre95}
Bregman, J. N., Schulman, E., \& Tomisaka, K. 1995, \apj, 439, 155
\bibitem[Burton et al.(1990)]{Bur90}
Burton, M. G., Hollenbach, D. J., Haas, M. R. \& Erickson, E. F. 1990, \apj, 355, 197
\bibitem[D\'esert et al.(1990)]{Des90}
     D\'esert, F. X. Boulanger, F., \& Puget J. L. 1990, \aap, 237, 215
\bibitem[Colbert et al.(1999)]{Col99}
Colbert, J. W., Malkan, M. A., Clegg, P. E., et al. 1999, \apj, 511, 721
\bibitem[Davies et al.(1998)]{Dav98}
     Davies, J. I., Alton, P., Bianchi, S., \& Trewhella, M. 1998, MNRAS, 300, 1006
\bibitem[Devine \& Bally(1999)]{Dev99}
Devine, D. \& Bally, J. 1999, \apj, 510, 197
\bibitem[Draine \& Li(2007)]{Dra07}
     Draine, B. T. \& Li, A. 2007, \apj, 657, 810
\bibitem[Eales \& Edmunds(1996)]{Eal96}
     Eales, S. \& Edmunds, M. 1996, MNRAS, 280, 1167
\bibitem[Engelbracht et al.(2006)]{Eng06}
Engelbracht, C. W., Kundurthy, P., Gordon, K. D., et al. 2006, \apj, 642, L12
\bibitem[F\"orster Schreiber et al.(2003a)]{For03a}
F\"orster Schreiber, N. M., Genzel, R., Lutz, D., \& Sternberg, A. 2003a, \apj, 599, 193
\bibitem[F\"orster Schreiber et al.(2003b)]{For03b}
F\"orster Schreiber, N. M., Sauvage, M., Charmandaris, V., Laurent, O., Gallais, P., Mirabel, I. F., \& Vigroux, L. 2003b, \aap, 399, 833
\bibitem[Garc\'ia-Burillo et al.(2001)]{Gar01}
     Garc\'ia-Burillo, S., Mart\'in-Pintado, J., \& Fuente, A. 2001, \apj, 563, L27
\bibitem[Heckman et al.(2000)]{Hec00}
   Heckman, T. M., Lehnert, M. D., Strickland, D. K., \& Armus, L. 2000, \apjs, 129, 493
\bibitem[Heisler \& Ostriker(1988)]{Hei88}
     Heisler, J. \& Ostriker, J. P. 1988, \apj, 332, 543 
\bibitem[Hildebrand(1983)]{Hil83}
     Hildebrand, R. H. 1983, QJRAS, 24, 267
\bibitem[Hoopes et al.(2005)]{Hoo05}
     Hoopes, C. G., Heckman, T. M., Strickland, D. K., et al. 2005, \apj, 619, L99
\bibitem[Joblin et al.(1994)]{Job94}
     Joblin, C., D'Hendecourt, L., Leger, A., \& Defourneau, D. 1994, \aap, 281, 923
\bibitem[Kaneda et al.(2007)]{Kan07}
Kaneda, H., Kim, W.-J., Onaka, T., Wada, T., Ita, Y., Sakon, I., \& Takagi, T. 2007, \pasj, 59, S423 
\bibitem[Kaneda et al.(2008a)]{Kan08a}
Kaneda, H., Onaka, T., Sakon, I., Kitayama, T., Okada, Y., \& Suzuki, T. 2008a, \apj 684, 270
\bibitem[Kaneda et al.(2008b)]{Kan08b}
Kaneda, H., Suzuki, T., Onaka, T., Okada, Y., \& Sakon, I. 2008b, \pasj, 60, S467  
\bibitem[Kaneda et al.(2009a)]{Kan09a}
Kaneda, H., Koo, B.-C., Onaka, T., \& Takahashi, H. 2009a, Adv.Sp.Res., 44, 1038
\bibitem[Kaneda et al.(2009b)]{Kan09b}
Kaneda, H., Yamagishi, M., Suzuki, T., \& Onaka, T. 2009b, \apj, 698, L125
\bibitem[Karachentsev et al.(2002)]{Kar02}
Karachentsev, I. D., Dolphin, A. E., Geisler, D., et al. 2002, \aap, 383, 125
\bibitem[Kawada et al.(2007)]{Kaw07}
Kawada, M., Baba, H., Barthel, P. D., et al. 2007, \pasj, 59, 389
\bibitem[Lallement(2004)]{Lal04}
Lallement, R. 2004, \aap, 422, 391
\bibitem[Leeuw \& Robson(2009)]{Lee09}
Leeuw, L. L. \& Robson, E. I. 2009, \apj, 137, 517
\bibitem[Lehnert et al.(1999)]{Leh99}
Lehnert, M. D., Heckman, T. M., \& Weaver, K. A. 1999, \apj, 523, 575 
\bibitem[Lorente, et al.(2007)]{Lor07}
Lorente, R., Onaka, T., Ita, Y., Ohyama, Y., \& Pearson, C. P. 2007, AKARI IRC Data User Manual Version 1.2
\bibitem[Murakami et al.(2007)]{Mur07}
Murakami, H., Baba, H., Barthel, P., et al. 2007, \pasj, 59, 369
\bibitem[Ohyama et al.(2002)]{Ohy02}
Ohyama, Y., Taniguchi, Y., Iye, M., et al. 2002, \pasj, 54, 891
\bibitem[Onaka et al.(2007)]{Ona07}
Onaka, T., Matsuhara, H., Wada, T., et al. 2007, \pasj, 59, 4010
\bibitem[Radovich et al.(2001)]{Rad01}
     Radovich, M., Kahanp\"a\"a, J., \& Lemke, D. 2001, A\&A, 377, 73
\bibitem[Shen \& Lo(1995)]{She95}
Shen, J. \& Lo, K. Y. 1995, \apj, 445, L99 
\bibitem[Shirahata et al.(2009)]{Shi09} 
Shirahata, M., Matsuura, S., Hasegawa, S., et al. 2009, \pasj, 61, 737
\bibitem[Shopbell \& Bland-Hawthorn(1998)]{Sho98}
Shopbell, P. L. \& Bland-Hawthorn, J. 1998, \apj 493,129
\bibitem[Smith et al.(2007)]{Smi07}
Smith, J. D., Draine, B. T., Dale, D. A., et al. 2007, \aap, 656, 770 
\bibitem[Sodroski et al.(1994)]{Sod94}
Sodroski, T. J., Bennett, C., Boggess, N., et al. 1994, \apj, 428, 638
\bibitem[Sodroski et al.(1997)]{Sod97}
Sodroski, T. J., Odegard, N., Arendt, R. G., Dwek, E., Weiland, J. L., Hauser, M. G., \& Kelsall, T. 1997, \apj, 480, 173
\bibitem[Strickland et al.(1997)]{Str97}
Strickland, D. K., Ponman, T. J., \& Stevens, I. R. 1997, \aap, 320, 378
\bibitem[Strickland \& Stevens(2000)]{Str00}
Strickland, D. K. \& Stevens, I. R. 2000, MNRAS, 314, 511
\bibitem[Strickland et al.(2004)]{Str04}
Strickland, D. K., Heckman, T. M., Colbert, E. J. M., Hoopes, C. G., \& Weaver, K. A. 2004, \apjs, 151, 193
\bibitem[Sturm et al.(2000)]{Stu00}
Sturm, E., Lutz, D., Tran, D., et al. 2000, \aap, 358, 481
\bibitem[Telesco \& Harper(1980)]{Tel80}
Telesco, C. M. \& Harper, D. A. 1980, \apj 235, 392
\bibitem[Thuma et al.(2000)]{Thu00}
Thuma, G., Neininger, N., Klein, U., \& Wielebinski, R. 2000, \aap, 358, 65
\bibitem[Tsuru et al.(2007)]{Tsu07}
Tsuru, T. G., Ozawa, M., Hyodo, Y. et al. 2007, \pasj, 59, S269
\bibitem[Veilleux et al.(2009)]{Vei09}
Veilleux, S., Rupke, D. S. N., \& Swaters, R. 2009, \apj, 700, L149
\bibitem[Verdugo, et al.(2007)]{Ver07}
Verdugo, E., Yamamura, I., \& Pearson, C. P. 2007, AKARI FIS Data User Manual Version 1.2
\bibitem[Walter et al.(2002)]{Wal02}
Walter, F., Weiss, A., \& Scoville, N. 2002, \apj, 580, L21
\bibitem[Watson et al.(1984)]{Wat84}
Watson, M. G., Stanger, V., \& Griffiths, R. E. 1984, \apj, 286, 144
\bibitem[Werner et al.(2004)]{Wer04}
Werner, M. W., Uchida, K. I., Sellgren, K., et al. 2004, \apjs, 154, 309 
\bibitem[Xilouris et al.(2006)]{Xil06}
Xilouris, E., Alton, P., Alikakos, J., Xilouris, K., Boumis, P., \& Goudis, C. 2006, \apj, 651, L107
\bibitem[Yun et al.(1994)]{Yun94}
Yun, M. S., Ho, P. T. P., \& Lo, K. Y. 1994, \nat, 372, 530
\bibitem[Yun et al.(1993)]{Yun93}
Yun, M. S., Ho, P. T. P., \& Lo, K. Y. 1993, \apj, 411, L17
\end{thebibliography}
\end{document}